\documentclass[a4paper,cleveref, autoref, thm-restate]{lipics-v2021}

\nolinenumbers

\usepackage{amsfonts}
\usepackage{amsmath}
\usepackage{mathtools}
\usepackage{todonotes}
\usepackage{subcaption}
\usepackage{hyperref}
\usepackage{graphicx}
\graphicspath{{./figs/}}

\makeatletter
\newcommand\footnoteref[1]{\protected@xdef\@thefnmark{\ref{#1}}\@footnotemark}
\makeatother

\newcommand{\gtwisted}{generalized twisted}

\title{Characterizing and Recognizing Twistedness}

\author{Oswin Aichholzer}{Graz University of Technology, Austria}{oswin.aichholzer@tugraz.at}{https://orcid.org/0000-0002-2364-0583}{}
\author{Alfredo Garc{\'i}a}{Departamento de M\'etodos Estad\'\i sticos and IUMA, University of Zaragoza, Spain}{}{https://orcid.org/0000-0002-6519-1472}{}
\author{Javier Tejel}{Departamento de M\'etodos Estad\'\i sticos and IUMA, University of Zaragoza, Spain}{jtejel@unizar.es}{https://orcid.org/0000-0002-9543-7170}{Partially supported by project E41-23R, funded by Gobierno de Arag\'on, and project PID2023-150725NB-I00, funded by MICIU/AEI/10.13039/501100011033.}
\author{Birgit Vogtenhuber}{Graz University of Technology, Austria}{birgit.vogtenhuber@tugraz.at}{https://orcid.org/0000-0002-7166-4467}{}
\author{Alexandra Weinberger}{FernUniversit\"at in Hagen, Germany}{alexandra.weinberger@fernuni-hagen.de}{https://orcid.org/0000-0001-8553-6661}{}
\authorrunning{Aichholzer, Garc{\'i}a, Tejel, Vogtenhuber, Weinberger}
\Copyright{Oswin Aichholzer, Alfredo Garc{\'i}a, Javier Tejel, Birgit Vogtenhuber, Alexandra Weinberger}


\hideLIPIcs

\begin{document}

\maketitle

\begin{abstract}
In a simple drawing of a graph, any two edges intersect in at most one point (either a common endpoint or a proper crossing).
A simple drawing is \emph{generalized twisted} if it fulfills certain rather specific constraints on how the edges are drawn.
An abstract rotation system of a graph assigns to each vertex a cyclic order of its incident edges.
A realizable rotation system is one that admits a simple drawing such that at each vertex, the edges emanate in that cyclic order,
and a generalized twisted rotation system can be realized as a generalized twisted drawing.
Generalized twisted drawings have initially been introduced to obtain improved bounds on the size of plane substructures in any simple drawing of $K_n$.
They have since gained independent interest due to their surprising properties.
However, the definition of generalized twisted drawings is very geometric and drawing-specific.

In this paper, we develop characterizations of generalized twisted drawings that enable a purely combinatorial view on these drawings
and lead to efficient recognition algorithms. 
Concretely, we show that for any $n \geq 7$, an abstract rotation system of $K_n$ is generalized twisted if and only if all subrotation systems induced by five vertices are generalized twisted.
This implies a drawing-independent and concise characterization of generalized twistedness. Besides, the result yields a simple $O(n^5)$-time algorithm to decide whether an abstract rotation system is generalized twisted and sheds new light on the structural features of simple drawings.
We further develop a characterization via the rotations of a pair of vertices in a drawing, which we then use to derive an $O(n^2)$-time
algorithm to decide whether a realizable rotation system is generalized twisted.
\end{abstract}
%
\section{Introduction}\label{sec:intro}

\begin{figure}[thbp]
	\begin{minipage}[t]{.45\linewidth}
		\centering\includegraphics[page=4]{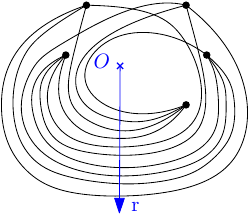}
	\end{minipage}
	\begin{minipage}[t]{.45\linewidth}
		\centering\includegraphics[page=10]{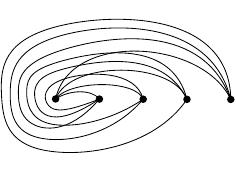}
	\end{minipage}
	\caption{Two realizations of the generalized twisted rotation system of $K_5$. \textsf{Left:} Generalized twisted drawing where the origin $O$ and ray $r$ are depicted. \textsf{Right:} The classical twisted drawing.}\label{fig:geom_gtwistedk6}
\end{figure}

\emph{Simple drawings} are drawings of graphs in which the vertices are drawn as distinct points in the plane, the edges are drawn as Jordan arcs connecting their end-vertices without passing through other vertices, and each pair of edges share at most one point (a proper crossing or a common endpoint).
We consider two simple drawings the same if they are strongly isomorphic, that is, if there is a homeomorphism of the plane transforming one drawing into the other.
In this work we focus on simple drawings of $K_n$. Unless explicitly stated otherwise, all considered drawings are of $K_n$ and simple.
A simple drawing $D$ is \emph{generalized twisted} if it is strongly isomorphic to a simple drawing in which there is a point~$O$ such that each ray emanating from~$O$ crosses each edge of $D$ at most once and there exists a ray $r$ emanating from~$O$ such that all edges of $D$ cross~$r$. For an example see Figure~\ref{fig:geom_gtwistedk6}(left), which has been reproduced from \cite{triangle_gtwisted_journal}.

Generalized twisted drawings have been used to find plane substructures in all simple drawings of $K_n$: The currently best bounds on the size of the largest plane matching and longest plane cycle and path in any simple drawing of $K_n$ have been obtained via generalized twisted drawings~\cite{plane_gtwisted_journal}. Moreover, generalized twisted drawings are the largest class of drawings for which it has been shown that each drawing has exactly $2n-4$ empty triangles~\cite{triangle_gtwisted_conference,triangle_gtwisted_journal}, which is conjectured to be the minimum over all simple drawings~\cite{ahprsv-etgdc-15}. There are several more interesting properties of generalized twisted drawings that have been shown already~\cite{plane_gtwisted_socg,plane_gtwisted_journal,triangle_gtwisted_journal}. For example, all generalized twisted drawings contain plane cycles of length at least $n-1$~\cite{plane_gtwisted_journal}.

Despite the usefulness of generalized twisted drawings, their definition is very geometrical and drawing-specific.
In this paper, we obtain combinatorial characterizations that no longer depend on the specific drawing, which ease the work with them and also allow for efficient recognition.
Note that ad hoc it is not clear that such a combinatorial characterization needs to exist.
The study of special classes of simple drawings~\cite{shellable,gconvex_cycles,convex_holes,pseudocycle_more} %
and their combinatorial characterizations~\cite{pseudolinear_char1,gconvex_char,monotone_char,pseudocycle_flip,pseudolinear_char2} has a strong tradition in the research of simple drawings
and contributes significantly to the understanding %
simple drawings in general.

A widely used combinatorial abstraction of drawings are rotation systems.
The \emph{rotation of a vertex} in a drawing is the cyclic order of its incident edges.
The \emph{rotation system of a simple drawing} is the collection of the rotations of all vertices.
Rotation systems are especially useful
for describing simple drawings of complete graphs,
since two simple drawings of~$K_n$ have the same crossing edge pairs if and only if they have the same or inverse rotation system~\cite{triangleflip,gioan_conference,gioan_RS,RS_weak}.
Rotation systems  of graphs can also be considered without a concrete drawing.
An \emph{abstract rotation system} of a graph $G$ gives, for every vertex of $G$, %
a cyclic order of its adjacent vertices.
An abstract rotation system is called \emph{realizable} if there exists a simple drawing with that rotation system.
For abstract rotation systems of %
complete graphs,
realizability can be checked in polynomial time. %
Specifically, an abstract rotation system of  $K_n$ %
is realizable if and only if every subrotation system induced by five vertices is realizable. This follows from
a combinatorial characterization via subrotation systems of size six~\cite[Theorem 1.1.]{realizability}
together with
a complete enumeration of all realizable rotation systems of $K_n$ for $n\leq 9$~\cite{all_small_drawings}. 

In this work, we use rotation systems to characterize generalized twisted drawings.
An abstract rotation system of a graph is \emph{generalized twisted} %
if there is a generalized twisted drawing realizing this rotation system.
As our first main result, we show that generalized twisted rotation systems of $K_n$ can also be characterized via subrotation systems of size five.
The main benefit of this result is to have a drawing-independent (and concise) characterization of generalized twistedness.
The substructures to be considered are of constant size and straightforward to be checked, which improves the insight into generalized twisted drawings. %

\begin{theorem}\label{thm:T5}
	Let $R$ be an abstract rotation system of $K_n$ with $n \ge 7$. Then $R$ is generalized twisted if and only if every subrotation system induced by five vertices is generalized twisted.
\end{theorem}

This characterization of generalized twisted rotation systems nicely complements the characterization of all realizable simple drawings in general,
which can also be done by considering subrotation systems of size five~\cite{all_small_drawings,realizability}.
Per se, such characterizations of bounded complexity do not always need to exist.
For example, if the combinatorics of an (abstract) set of points is given by their triple orientations -- so-called abstract order types -- no simple characterization to decide their realizability as a point set in the plane can exist, as this decision problem is known to be $\exists\mathbb{R}$-complete; see the recent survey~\cite{existentialtheoryreals} for details.

\begin{figure}[!tb]
	\centering
	\includegraphics[scale=0.75]{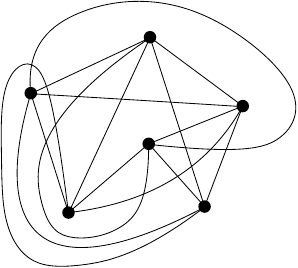}
	\caption{A simple drawing of $K_6$ whose rotation system is not generalized twisted, even though all subrotation systems on five vertices are.}
	\label{fig:special_k6}
\end{figure}

The main part of Theorem~\ref{thm:T5} %
is the ``if'' part. The ``only-if'' part follows from the fact that subdrawings of generalized twisted drawings are generalized twisted by definition.
As mentioned in the article introducing generalized twisted drawings~\cite{plane_gtwisted_socg},
there is exactly one generalized twisted rotation system $T_5$ of~$K_5$, namely, the one of the \emph{twisted} drawing~\cite{harborth,pachetal} of~$K_5$, which gave rise to the name ``generalized twisted''; see Figure~\ref{fig:geom_gtwistedk6}(right) for an illustration as in~\cite{pachetal}. %
We remark that Theorem~\ref{thm:T5} cannot be extended to $n\le6$.
Figure~\ref{fig:special_k6} depicts a drawing from~\cite{plane_gtwisted_socg}, where it has been shown that the rotation system of this drawing is not generalized twisted, although all its subrotation systems are (generalized) twisted.

\begin{figure}[!ht]
	\centering
	\includegraphics[scale=0.85,page=17]{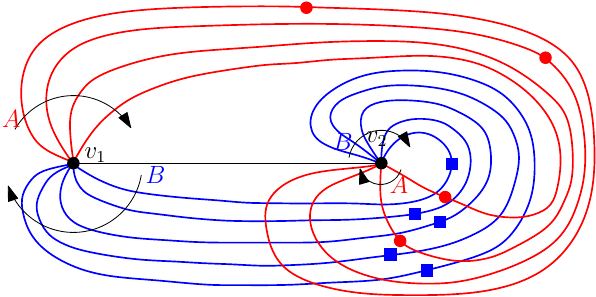}
	\caption{Illustration of Theorem~\ref{the:bipartition}: Possible rotations of $v_1$ and $v_2$. The vertices of partition set~$A$ are drawn as red disks and the vertices of partition set $B$ as blue squares.}
	\label{fig:bipartition_thm}
\end{figure}
Our second main result is a
characterization of generalized twisted drawings via rotation systems and crossings as stated in Theorem~\ref{the:bipartition} below and depicted on an example in Figure~\ref{fig:bipartition_thm}.

\begin{figure}[!thbp]
		\centering\includegraphics[page=5]{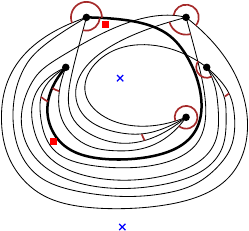}
	\caption{A realization of the generalized twisted rotation system $T_5$ of $K_5$. The pairs of antipodal vi-cells are marked with red squares and blue crosses, the maximum crossing edge is bold, and empty triangles are indicated by (brown) circular arcs.}\label{sfig:lemmaw2}
\end{figure}

\begin{theorem}\label{the:bipartition}
	Let $R$ be the rotation system of a simple drawing $D$ of $K_n$ and let $V$ be the set of vertices of $D$.
	Then $R$ is generalized twisted if and only if there exist two vertices $v_1$ and $v_2$ in $V$ and a bipartition $A\cup B$ of the vertices in $V\setminus \{v_1,v_2\}$,
	where some of $A$ or $B$ can be empty, such that:
	\begin{enumerate}
		\item[(i)] For every pair of vertices $a$ and $a'$ in $A$, the edge $aa'$ crosses~$v_1v_2$.
		\item[(ii)] For every pair of vertices $b$ and $b'$ in $B$, the edge $bb'$ crosses~$v_1v_2$.
		\item[(iii)] For every vertex $a\in A$ and every vertex $b\in B$, the edge $ab$ does not cross~$v_1v_2$.
		\item[(iv)] Beginning at~$v_2$, in the rotation of~$v_1$, all vertices in $B$ appear before all vertices in~$A$.
		\item[(v)] Beginning at~$v_1$, in the rotation of~$v_2$, all vertices in $B$ appear before all vertices in~$A$.
	\end{enumerate}
\end{theorem}

The main implication of Theorem~\ref{the:bipartition} is that it leads to fast algorithms for recognizing generalized twisted rotation systems.
If it is already known that a rotation system is realizable (which needs $O(n^5)$ time to be checked~\cite{all_small_drawings,realizability}), we can even use it to obtain a quadratic time recognition algorithm.

\begin{theorem}\label{thm:algo}
	Let $R$ be an abstract rotation system of~$K_n$. Then it can be decided in $O(n^5)$ time whether $R$ is generalized twisted. If $R$ is known to be realizable, then deciding whether it is also generalized twisted can be done in $O(n^2)$ time.
\end{theorem}

\noindent \textsl{\textbf{Outline.}}
We start by introducing some notation and showing first properties in Section~\ref{sec:pre}. Sections~\ref{sec:5} and~\ref{sec:bipartition} are devoted to proving the two characterizations, Theorem~\ref{thm:T5} and Theorem~\ref{the:bipartition}, respectively. In Section~\ref{sec:algo}, we prove Theorem~\ref{thm:algo} by providing the algorithms. We conclude in Section~\ref{sec:conclusion} with some open questions.

\noindent \textit{The majority of the proofs are deferred to the appendix.}

\section{Preliminaries}\label{sec:pre}
Given a simple drawing $D$, the \emph{star at v}, denoted by $S(v)$, is the set of edges incident to $v$ in~$D$.
For a set $V$ of vertices of $D$, we denote by $D \setminus V$ the drawing $D$ without the vertices in $V$ and without all edges incident to a vertex in $V$.
A \emph{triangle} $\Delta$ in $D$ is the simple closed curve formed by three vertices and the three edges connecting them.
Every triangle partitions the plane and hence $D$ into two regions which we call the \emph{sides} of $\Delta$.
The whole drawing $D$ partitions the plane into regions called \emph{cells}.

In our proofs, we heavily rely on antipodal vi-cells, empty star triangles, and maximum crossing edges, all of which we describe in the following paragraphs.%

\medskip

\noindent \textsl{\textbf{Antipodal vi-cells.}}
A cell %
that is incident to a vertex (that is, has the vertex on its boundary) is called vertex-incident cell or \emph{vi-cell}.
We observe that 
if there is a vi-cell at $v$ containing the initial parts of $vu$ and $vw$ in one drawing of a rotation system, then 
$vu$ and $vw$ are neighboring in the rotation of $v$. 
Since the latter property depends only on the rotation system of the drawing, 
there is a vi-cell at $v$ in every drawing of this rotation system with initial parts of $vu$ and $vw$ on the boundary.
Note that the remainder of the boundary of the vi-cell %
is not necessarily identical across different drawings of the rotation system.
Two cells are \emph{antipodal} if for every triangle, the two cells are on different sides.
A vi-cell that is antipodal to some other vi-cell %
is called a \emph{via-cell}.
(In Figure~\ref{sfig:lemmaw2}, the two pairs of via-cells are marked.)
Since the set of crossing edge pairs determines the rotation system and vice versa~\cite{gioan_RS,RS_weak},
also the existence of via-cells is a property of the rotation system.
A vertex incident to a via-cell is called a \emph{via-vertex}.

We will need the following known theorem~\cite[Theorem 15]{plane_gtwisted_socg}%
\footnote{\label{footnoteRS} The original formulation uses drawings and weak isomorphism. Our formulation is equivalent due to the equivalence of rotation systems and weak isomorphism~\cite{gioan_RS,RS_weak}.}.

\begin{theorem}[\cite{plane_gtwisted_socg}]\label{thm:antipodal_gtwisted}
	The rotation system of a simple drawing $D$ of $K_n$ is generalized twisted if and only if $D$ contains a pair of antipodal vi-cells.
\end{theorem}

Note that the characterization in Theorem~\ref{thm:antipodal_gtwisted} implies that a rotation system is generalized twisted if and only if \emph{all its realizations as simple drawings} contain a pair of antipodal vi-cells. %

\medskip

\noindent \textsl{\textbf{Empty star triangles.}}
A triangle~$\Delta$ with vertices $x, y$ and $z$ is a \emph{star triangle} at $x$ if the star $S(x)$ 
does not cross the edge $yz$;
it is \emph{empty} if one side of $\Delta$
contains no vertices of $D\setminus \Delta$ (and hence the other side contains all of them).
(In Figure~\ref{sfig:lemmaw2}, each circular arc between two edges at a vertex indicates that the triangle containing those two edges is an empty star triangle at that vertex.)
We call two star triangles at $x$ \emph{adjacent} if they are both incident to some edge $xy$ and span disjoint angles at~$x$. %
We will use the following results on empty star triangles in simple and generalized twisted drawings. %
\begin{lemma}[\cite{ahprsv-etgdc-15}]\label{lem:gen_startriangles}
Let $v$ be a vertex of a simple drawing $D$ of $K_n$. Then $D$ contains at least two empty star triangles at $v$.
If $vxy$ is a star triangle at $v$ then it is empty if and only if the vertices $x$ and $y$ are consecutive in the rotation of $v$.
\end{lemma}
\begin{lemma}[\cite{triangle_gtwisted_journal}]\label{lem:gtwisted_startriangles}
	Let $D$ be a generalized twisted drawing of $K_n$ and let $v$ be a vertex of~$D$. Then there are exactly two empty star triangles at~$v$. If $(C_1,C_2)$ is a pair of antipodal vi-cells in $D$, then one of the two empty star triangles at $v$ contains $C_1$ on the empty side and the other triangle contains $C_2$ on the empty side.
\end{lemma}

\medskip

\noindent \textsl{\textbf{Maximum crossing edges.}}
An edge $v_1v_2$ is \emph{maximum crossing} if $v_1v_2$ crosses every edge not in $S(v_1) \cup S(v_2)$. (For example, the bold edge in Figure~\ref{sfig:lemmaw2} is maximum crossing.) Any simple drawing has at most one maximum crossing edge, as the following lemma states.
\begin{lemma}\label{lem:only_one_maxcross}
	Any simple drawing of $K_n$ with $n \geq 5$ has at most one maximum crossing edge.
\end{lemma}

Lemma~\ref{lem:only_one_maxcross}  holds for $n=5$ because none of the five simple drawings of $K_5$ (up to strong isomorphism)~\cite{ahprsv-etgdc-15} contains more than one maximum crossing edge. Using this fact, the lemma can be proven for $n>5$ by contradiction, assuming that a simple drawing contains several maximum crossing edges.

\begin{figure}[!htb]
	\centering
	\includegraphics[scale=0.8,page=1]{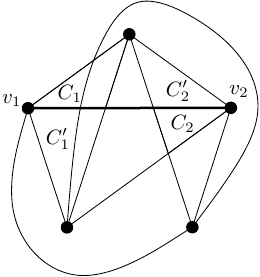}
	\caption{A maximum crossing edge $v_1v_2$ (in bold) and the two pairs, $(C_1,C_2)$ and $(C'_1,C'_2)$, of antipodal vi-cells.}
	\label{fig:Lemma8}
\end{figure}

We next observe in Lemma~\ref{lem:maxcross_then_gtwisted} that maximum crossing edges imply generalized twisted rotation systems. Figure~\ref{fig:Lemma8} illustrates the lemma.%
\begin{lemma}\label{lem:maxcross_then_gtwisted}
	Let $D$ be a simple drawing of $K_n$ that has a maximum crossing edge, $v_1v_2$.  Let $C_1$ and $C'_1$ be the two cells incident to $v_1$ along the edge $v_1v_2$ and let $C_2$ and $C'_2$ be the two cells incident to $v_2$ along the edge $v_1v_2$ such that when going along the edge from $v_1$ to~$v_2$, the cell $C_1$ is on the other side of $v_1v_2$ than the cell $C_2$ (and consequently $C'_1$ is on the other side of $v_1v_2$ than $C'_2$). Then $D$ is generalized twisted and $(C_1,C_2)$ and $(C'_1,C'_2)$ are two pairs of antipodal vi-cells.
\end{lemma}

Lemma~\ref{lem:maxcross_then_gtwisted} can be shown %
by taking an arbitrary triangle of the generalized twisted drawing and considering where the cells adjacent to the edge $v_1v_2$ and its endpoints must lie with respect to this triangle in order to allow the edge $v_1v_2$ to cross all non-adjacent edges.

\section{Characterization via subrotation systems induced by five vertices}\label{sec:5}

In this section, we prove Theorem~\ref{thm:T5} by induction. To this end, we use the following theorem, whose proof we sketch in Section~\ref{ssec:step}.

\begin{theorem}\label{thm:step}
	Let $R$ be an abstract rotation system of $K_n$ with $n \geq 12$ such that every induced subrotation system on at most $n-1$ vertices is generalized twisted. Then $R$ is generalized twisted.
\end{theorem}

\begin{proof}[Proof of Theorem~\ref{thm:T5}]
By definition, if $R$ is generalized twisted, then any subrotation system induced by five vertices is generalized twisted.
We prove the converse:
if every subrotation system induced by five vertices is generalized twisted, then $R$ is generalized twisted.

For $7\le n\le 11$, the theorem is verified by computer. Following the lines of~\cite{extra_material,plane_gtwisted_journal}, we compute all generalized twisted rotation systems of $K_n$ for $n$ up to $15$ in the following way. First, we use the method of~\cite{all_small_drawings} to generate all realizable rotation systems, but discard those with subrotation systems induced by five vertices different to $T_5$, thus computing all candidate rotation systems for generalized twisted drawings.
In each candidate rotation system, we then search for a pair of antipodal vi-cells.
By Theorem~\ref{thm:antipodal_gtwisted}, this identifies all generalized twisted rotation systems.
Any rotation system contradicting Theorem~\ref{thm:T5} does not have antipodal vi-cells. The computational results give only one %
rotation system of size $n=6$ %
(drawn in Figure~\ref{fig:special_k6})
that is not generalized twisted. 

For $n\ge 12$, we use induction. %
Every induced subrotation system of $R$ on at most $n-1$ vertices is generalized twisted by induction. Thus, $R$ is generalized twisted by Theorem~\ref{thm:step}.
\end{proof}

\subsection{Proof sketch of Theorem~\ref{thm:step}}\label{ssec:step}

Let $R$ be as in Theorem~\ref{thm:step} and let $D$ be a realization of $R$, which exists since any
abstract rotation system of $K_n$ is realizable if all subrotation systems induced by five vertices
are~\cite{all_small_drawings,realizability}. 
Since each induced and proper 
subdrawing of $D$ has a generalized twisted rotation system,  it 
also has a pair of antipodal vi-cells by Theorem~\ref{thm:antipodal_gtwisted}. We use this to show that also $D$ has such a pair 
and thus, by Theorem~\ref{thm:antipodal_gtwisted}, $R$ is generalized twisted.
The idea is to locate antipodal vi-cells of $D$ using those in its induced subdrawings and empty triangles.
The following lemma can be shown using Lemma~\ref{lem:gen_startriangles} (for a lower bound) and Lemma~\ref{lem:gtwisted_startriangles} (for an upper bound).

\begin{lemma}\label{lem:exactly_two}
Let $D$ be a drawing of $K_n$ with $n\ge 5$ such that every induced and proper subdrawing has a {\gtwisted} rotation system. Then at any vertex of $D$, there are exactly two empty star triangles.
\end{lemma}

We separately consider the case where $D$ contains a vertex $u$ with two adjacent empty star triangles at $u$ %
and the case that there is no such vertex.

\bigskip
\noindent \textbf{Case 1: Two adjacent empty star triangles at a vertex $u$} \\ %
This is the easy case. We first show the following lemma on generalized twisted rotation systems of~$K_n$.
Its proof is based on considering two adjacent empty star triangles at a vertex $v_1$ %
and analyzing all possibilities of how to extend these triangles to a drawing of $T_5$ and then to a drawing of $K_6$ with a generalized twisted rotation system.

\begin{lemma}\label{lem:adjacent_then_maxcross}
	Let $v_1$ and $v_2$ be two vertices of a simple drawing of $K_{n}$ with generalized twisted rotation system such that the two empty star triangles at vertex $v_1$ are sharing the edge~$v_1v_2$.
Then the following properties hold.
\begin{enumerate}
	\item\label{lem:adjacent_then_maxcross_i} The edge $v_1v_2$ is maximum crossing.
	\item\label{lem:adjacent_then_maxcross_ii} The two empty star triangles at $v_1$ are also empty star triangles at $v_2$.	\item\label{lem:adjacent_then_maxcross_iii} For each of the two empty star triangles at $v_1$, the vertices $v_1$ and $v_2$ are adjacent in the rotation of the third vertex of the triangle in such a way that all edges emanate from the third vertex in the empty side of the triangle.
	\end{enumerate}
\end{lemma}

Then we consider the drawing $D$ and the two adjacent empty star triangles at $u$ in $D$.
As every proper subdrawing of $D$ has a generalized twisted rotation system, Lemma~\ref{lem:adjacent_then_maxcross} implies that
the edge shared by the two adjacent empty star triangles at $u$ is maximum crossing. Thus, the rotation system of $D$ is generalized twisted by Lemma~\ref{lem:maxcross_then_gtwisted}.

\bigskip
\noindent \textbf{Case 2: No vertex with two adjacent empty star triangles.} \\
This is the involved case.
Let $v_1$ be an arbitrary vertex of $D$ and let $\Delta_1=v_1u_1u'_1$ and $\Delta_2=v_1u_2u'_2$ be the (nonadjacent) empty star triangles at $v_1$.
Our goal is to show that using via-cells of some induced and proper subdrawings of $D$, we can find a pair $(C_1, C_2)$ of vi-cells in $D$ such that $C_1$ lies in $\Delta_1$ and $C_2$ in $\Delta_2$, and prove that $C_1$ and $C_2$ are antipodal.

\begin{figure}[!htb]
	\centering
	\includegraphics[scale=0.8,page=1]{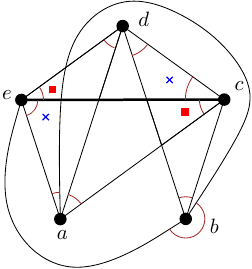}
	\caption{A drawing of the unique generalized twisted rotation system $T_5$ of $K_5$, which is strongly isomorphic to the one in Figure~\ref{fig:geom_gtwistedk6}, but redrawn to make empty star triangles and the relevant cells more visible. Empty star triangles at a vertex are marked with brown arcs at the angle that spans the triangle.  The two pairs of antipodal vi-cells are marked with red squares and blue crosses.}
	\label{fig:triangle_vi}
\end{figure}

Let $V_0$ be the set of vertices of $\Delta_1$ and $\Delta_2$, and let $D_0$ be the subdrawing induced by~$V_0$. %
Since the rotation system of $D_0$ is generalized twisted and has only five vertices, we have all the information about that drawing.
This includes in particular the  maximum crossing edge of~$D_0$, which we denote by $m$,
the two pairs of antipodal vi-cells of $D_0$, and the empty star triangles.
As $\Delta_1$ and $\Delta_2$ are not adjacent,
$v_1$ is one of the vertices $\{a,b,d\}$ in Figure~\ref{fig:triangle_vi}.

\begin{figure}[!htb]
\centering
	\includegraphics[page=5]{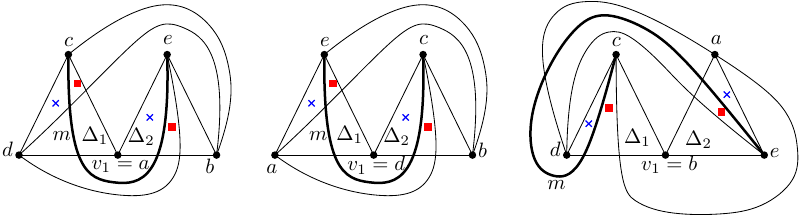}
	\caption{Three drawings realizing $T_5$ in such a way that $v_1$ is drawn in the center. The underlying unlabeled drawings are strongly isomorphic. The labels correspond to those in Figure~\ref{fig:triangle_vi} with $v_1=a$ (left), $v_1=d$ (middle), or $v_1=b$ (right). The maximum crossing edge $m$ is drawn bold, the two pairs of antipodal vi-cells are marked with red squares and blue crosses.}\label{fig:case2}
\end{figure}

Observe that the vertices $a$ and $d$ behave analogously to each other while vertex $b$ behaves differently (see Figure~\ref{fig:case2}).
We argue via the triangles $\Delta_1$ and $\Delta_2$ separately but simultaneously, because we will consider only one triangle but require that the other triangle has the same properties.
We sketch the case where $v_1$ is $a$ or $d$ and focus mostly on $\Delta_1$.
The properties for $\Delta_2$ can be shown analogously and the case of $v_1$ behaving like $b$ then follows immediately.

For identifying all pairs of antipodal vi-cells in subdrawings of $D$, we state in Lemma~\ref{lem:two_pairs_adjacent} properties of simple drawings with more than one such pair. %
\begin{lemma}\label{lem:two_pairs_adjacent}
	Let $D$ be a simple drawing of $K_n$ with different antipodal vi-cell pairs $(C_1,C_2)$ and~$(C'_1,C'_2)$. Then
	\begin{enumerate}
		\itemsep 0pt
		\item\label{slem:p1} For one of the vertices on the boundary of a cell in $\{C_1,C_2,C'_1,C'_2\}$, there are two empty star triangles at that vertex that share an edge.
		\item\label{slem:p2} There is a maximum crossing edge which is the shared edge of the two empty star triangles at one of the endpoints of the edge.
		\item\label{slem:p3} The via-cells are adjacent to the maximum crossing edge (as 
			described in Lemma~\ref{lem:maxcross_then_gtwisted}).
		\item\label{slem:p4} $D$ has only those two (and no more) pairs of antipodal vi-cells.
	\end{enumerate}
\end{lemma}	

\begin{proof}[Proof Sketch]

The proof of (\ref{slem:p1}) is by contradiction. Assuming that the two empty star triangles at any vertex on the boundary of a cell in $\{C_1,C_2,C'_1,C'_2\}$ are not adjacent, we can find an induced subdrawing on 5 or 6 vertices that is not generalized twisted, a contradiction. (\ref{slem:p2}) and (\ref{slem:p3}) follow from Lemma~\ref{lem:adjacent_then_maxcross}, as by (\ref{slem:p1}) there are two empty star triangles that are adjacent. (\ref{slem:p4})~ follows from the fact that if there is a third pair of antipodal vi-cells, then $D$ would contain two maximum crossing edges, contradicting  Lemma~\ref{lem:only_one_maxcross}.
\end{proof}

Let $V'$ be a nonempty subset of $V\setminus V_0$. By Lemma~\ref{lem:two_pairs_adjacent} and Lemma~\ref{lem:gtwisted_startriangles}, one of the following holds for the
subdrawing $D\setminus{V'}$ of $D$:
\begin{enumerate}
\item $D\setminus{V'}$ contains exactly one pair of antipodal vi-cells. One of those cells is in the empty side of triangle~$\Delta_1$ and the other one is in the empty side of triangle $\Delta_2$.
\item $D\setminus{V'}$ contains exactly two pairs of antipodal vi-cells. %
	For each pair, one cell %
	is in the empty side of~$\Delta_1$ %
	and the other one is in the empty side of %
	$\Delta_2$.
	Further, for each of %
	$\Delta_1$ and $\Delta_2$, %
		the two vi-cells in its empty side are adjacent along the maximum crossed edge of $D\setminus{V'}$ and incident to the endpoints of this edge. %
\end{enumerate}

The next lemma states the possibilities of how \emph{arbitrary antipodal vi-cell pairs} of two subdrawings $D\!\setminus\! \{ v\}$ and $D\!\setminus \!\{ w\}$ of $D$ can be contained in \emph{all pairs} of the subdrawing $D\!\setminus\! \{ v,w\}$.
The proof uses observations on the containment of cells of a drawing $D'$ in cells of its subdrawings and consequences of two cells being antipodal.
We denote by $(C_1^{v_1v_2\ldots}\!,C_2^{v_1v_2\ldots})$ an antipodal vi-cell pair in the drawing~$D\!\setminus\! \{ v_1, v_2, \ldots\}$.

\begin{figure}[!htb]
	\centering
	\includegraphics[scale=0.7,page=2]{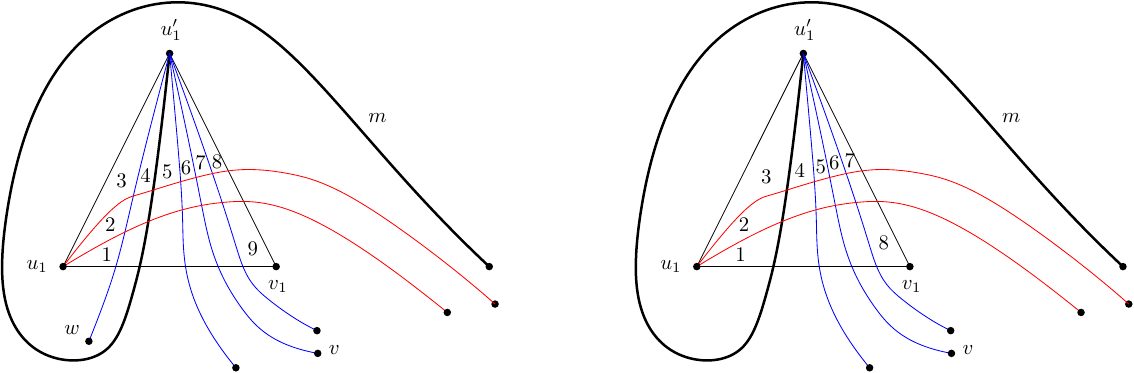}
	\caption{The grid cells of $D$ and $D\! \setminus\! {\{ w\} }$ inside triangle $\Delta_1$.}
	\label{GridCells}
\end{figure}

\begin{lemma}\label{lem:included_cells}
    For any pair $(C_1^{v}, C_2^{v})$ of antipodal vi-cells in $D\!\setminus\!\{v\}$ and any pair $(C_1^{w}, C_2^{w})$ of antipodal vi-cells in $D\!\setminus\!\{w\}$, then
\begin{enumerate}
	\item If $D\! \setminus\! \{v,w\}$ has %
	\textbf{only one antipodal vi-cell pair},
	$(C_1^{vw}\!,C_2^{vw})$,
	then $(C_1^{v}, C_2^{v})$ and $(C_1^{w}, C_2^{w})$ both are contained in $(C_1^{vw}, C_2^{vw})$.
	\item If $D\!\setminus\! {\{ v,w\} }$ has %
	\textbf{two antipodal vi-cell pairs}
	$(C_1^{vw}, C_2^{vw})$ and $({C_1'}^{vw}, {C_2'}^{vw})$, then either $(C_1^{v}, C_2^{v})$ and $(C_1^{w}, C_2^{w})$ both are contained in one of $(C_1^{vw}, C_2^{vw})$ and $({C_1'}^{vw}, {C_2'}^{vw})$, or $(C_1^{v}, C_2^{v})$ is contained in one of $(C_1^{vw}, C_2^{vw})$ and $({C_1'}^{vw}, {C_2'}^{vw})$, and $(C_1^{w}, C_2^{w})$ in the other.
\end{enumerate}
\end{lemma}

We identify grid structures in the empty sides of $\Delta_1$ and $\Delta_2$ in $D$ that we will work with to locate candidates for antipodal vi-cells of~$D$.
As $\Delta_1$ is an empty star triangle at $v_1$ and as $D$ is a simple drawing, no edge incident to $v_1$ can cross an edge of $\Delta_1$.
However, every generalized twisted drawing of~$K_4$ contains a crossing~\cite{plane_gtwisted_journal}.
Hence, for every vertex $v \notin \Delta_1$, exactly one of $vu_1$ or $vu'_1$ crosses $\Delta_1$ (on $v_1u'_1$ or $v_1u_1$, respectively).
We denote this edge by~$r_v$.
The set of edges $r_v$ for every $v$ 
form a grid in the empty side of~$\Delta_1$ in $D$.
We denote the cells of this grid that have some vertex (of $\Delta_1$) on the boundary as \emph{grid cells}; see Figure~\ref{GridCells}.
Grid cells in~$\Delta_2$ are defined analogously.
Note that grid cells may consist of several cells of~$D$, as edges of $D$ can cross through $\Delta_1$ and/or $\Delta_2$.

Next, we obtain an order on the grid cells of $\Delta_1$ (and analogously on $\Delta_2)$.
Consider the dual graph of the grid cells (where the vertices are the grid cells and two vertices share an edge if the corresponding grid cells are adjacent).
This dual graph
is either a path or a path plus an isolated vertex %
(e.g. grid cell $9$ in Figure~\ref{GridCells} left).
We label the grid cells $G_1,\ldots ,G_l$, along this dual path, beginning with the grid cell with vertex $u_1$ and adjacent to edge $u_1v_1$, and finishing with the cell that has $v_1$ on its boundary; see Figure \ref{GridCells} left for an example. %
The \emph{distance} between two grid cells $G_i,G_j$ is the number of edges $r_v$ separating $G_i$ and $G_j$. %
If edge $r_v$ is incident to both $G_i$ and $G_{i+1}$, we say that $G_i$ and $G_{i+1}$ are \emph{glued along}~$r_v$.
We denote the two grid cells that are glued along the edge $m$ (which is the maximum crossed edge of $D_0$) by $G_c$ and $G_{c+1}$ (cells $4,5$ in Figure \ref{GridCells} left).
We say that grid cell $G_i$ is \emph{placed before} $m$ if $i\leq c$; and \emph{placed after} $m$ otherwise.

When removing a vertex $v \in V\setminus V_0$, we obtain a grid in $D\! \setminus\! {\{ v\} }$ whose only difference to the grid in $D$ is that two grid cells of $D$ that are adjacent at the edge $r_v$ become one grid cell in $D\! \setminus\! {\{ v\} }$, and that some grid cells may enlarge their size, depending on the removed vertex; see Figure~\ref{GridCells} right for an illustration. %
For each via-cell $C^{v}$ of $D\! \setminus\! \{ v\}$ %
in the empty side of~$\Delta_1$, we denote by $G^{v}$ the grid cell of $D \!\setminus\!\{ v\} $ containing~$C^{v}$. %
Observe that $G^{v}$ is either an (enlarged) grid cell of $D$ or two such grid cells glued along an edge $r_v$.
With the following lemma, we obtain that the grid cells containing via-cells of the subdrawings have common intersections in the grid cells of $D$.

\begin{lemma}\label{lem:Claim1}
Let $w_1, w_2, \ldots , w_s$ be the set of vertices in $V\!\setminus \! V_0$ such that for every $w_i$, the grid cell $G^{w_i}$ in $D\!\setminus\! \{w_i\}$ is placed before $m$. Let $x_1, x_2, \ldots , x_t$ be the set of vertices in $V\!\setminus \! V_0$ such that for every $x_i$, the grid cell $G^{x_i}$ in $D\!\setminus\! \{x_i\}$ is placed after $m$. Then
\begin{enumerate}
	\itemsep 0pt
	\item \label{p:claim1_1} If $s\ge 2$, the common intersection of $G^{w_1}, G^{w_2}, \ldots , G^{w_s}$ is a grid cell %
	of $D$ placed before~$m$.
	\item \label{p:claim1_2} If $t\ge 2$, the common intersection of $G^{x_1}, G^{x_2}, \ldots , G^{x_t}$ is a grid cell %
	of $D$ placed after~$m$.
	\item \label{p:claim1_3} It $s\ge 2$ and $t\ge 2$, then 
	the two obtained grid cells of $D$ are $G_c$ and $G_{c+1}$, respectively. 	
\end{enumerate}
\end{lemma}

\begin{proof}[Proof Sketch]

(1) Given $w_i$ and $w_j$, we show that $G^{w_i}\cap G^{w_j}$ is a grid cell of $D$ by analyzing the intersections among $G^{w_i}, G^{w_j}$ and $G^{w_iw_j}$, for the different possibilities of the positions of $r_{w_i}$ and $r_{w_j}$. Then we show by contradiction that $G^{w_1},$ $G^{w_2}, \ldots , G^{w_s}$ have a common intersection, which is a grid cell of $D$ placed before~$m$. Assuming that $G^{w_i}\cap G^{w_j} \ne G^{w_j}\cap G^{w_k}$, we show that $G^{w_i}$ and $G^{w_k}$ cannot be contained in $G^{w_iw_k}$, a contradiction. (2) is proved in a similar way. To prove (3), we show by contradiction that $G_c$ (respectively $G_{c+1}$) is contained in $G^{w_i}$ (respectively $G^{x_i}$) for any $i$, so $G_c$ and $G_{c+1}$ must be the common grid cells.
\end{proof}

Next we show that with enough vertices in total, we also 
obtain vi-cells of $D$ in the (shared) grid cells we found in~$D$.

\begin{lemma}\label{lem:Claim2}
	Suppose that for the via-cells $C^{w_1}, C^{w_2}, \ldots, C^{w_s}$ with $s\ge 2$, the corresponding grid cells $G^{w_1}, G^{w_2}, \ldots, G^{w_s}$ in $D\! \setminus\!\{ w_1\}, D\! \setminus\!\{ w_2\}, \ldots, D\! \setminus\!\{ w_s\}$, respectively, contain a common grid cell $G$ of $D$. Then $C^{w_1}, C^{w_2}, \ldots, C^{w_s}$ also contain a common vi-cell $C$ of $D$.
\end{lemma}

\begin{proof}[Proof Sketch]
We first prove that if $G^{w_i}$ and $G^{w_j}$ contain a common grid cell $G$, then the interior of $C^{w_i}$ (or $C^{w_j}$) and the interior of $G$ are not disjoint. Using this result, if $G$ has only one vertex on its boundary, then the lemma trivially holds. If $G$ contains two vertices on its boundary, we partition $C^{w_1}, C^{w_2}, \ldots, C^{w_s}$ into three sets $S_1, S_2$, and $S_3$, where $S_1$ contains the cells $C^{w_i}$ that are incident to only one of the vertices, $S_2$ contains the cells $C^{w_i}$ that are incident to only the other vertex, and $S_3$ contains the cells $C^{w_i}$ that are incident to both vertices. We then prove that one of $S_1$ and $S_2$ must be empty, so $C^{w_1}, C^{w_2}, \ldots, C^{w_s}$ must all be incident to the same vertex of $G$ and share a common vi-cell $C$ of $D$.
\end{proof}

The following lemma shows that $D$ contains a pair of antipodal vi-cells.

\begin{lemma}\label{claim:proofend}%
For $n\ge 12$, there exists a pair $(C,C')$ of antipodal vi-cells in $D$. %
\end{lemma}

\begin{proof}[Proof Sketch]
For $i=1, \ldots , 7$, let $w_i$ be a vertex not in $V_0$ and let $(C^{w_i}_1, C^{w_i}_2)$ be a pair of antipodal vi-cells in the subdrawing $D\! \setminus\!\{ w_i\}$. %
For any pair of antipodal vi-cells of $V_0$, one of the cells is in $\Delta_1$ and the other in $\Delta_2$. Hence,  by Lemma~\ref{lem:included_cells}, $C_1^{w_i}$ must lie in $\Delta_1$ and $C_2^{w_i}$ in $\Delta_2$ for any $i$. Using Lemmas~\ref{lem:Claim1} and~\ref{lem:Claim2}, we show that at least four of the vi-cells $C_1^{w_i}$ share
a common vi-cell $C$ in $\Delta_1$ %
and a common vi-cell $C'$ in $\Delta_2$ for the corresponding vi-cells $C_2^{w_i}$.
By the antipodality of the pairs $(C^{w_i}_1, C^{w_i}_2)$ that contain $(C,C')$, we prove the antipodality of $(C,C')$ by showing for all triangles of $D$ that $C$ and~$C'$ are on different sides.
\end{proof}

\section{Characterization via bipartition}\label{sec:bipartition}
This section is devoted to the proof of Theorem~\ref{the:bipartition}.
Before starting with the proof, we introduce 
the concept of compatibility of vertices.
Given a realizable rotation system of~$K_n$, let $R_{v_1}= \{v_2=w_1, w_2, \ldots,  w_{n-1}\}$ and $R_{v_2} = \{v_1=t_1, t_2, \ldots,  t_{n-1}\}$ be the (clockwise) rotations of $v_1$ and $v_2$, respectively. We say that $v_1$ and $v_2$ are {\em compatible} (or that $v_1$ is compatible with $v_2$) if $\{v_1, t_2, \ldots,  t_{n-1}\}= \{v_1, w_i, w_{i-1}, \ldots, w_2,w_{n-1}, w_{n-2}, \ldots, w_{i+1}\}$, for some $i \in \{1, 2, \ldots, n\}$.
Note that if $v_1$ and $v_2$ are compatible and $i$ is 1 or $n$, then $\{v_1, t_2, \ldots,  t_{n-1}\}= \{v_1,w_{n-1}, w_{n-2}, \ldots, w_{2}\}$.

Next we consider pairs of antipodal vi-cells in simple drawings (and thus in generalized twisted drawings) and show that each such pair contains a pair of compatible vertices.
For any antipodal vi-cell pair $(C_1, C_2)$ in a simple drawing of $K_n$, there exist two vertices $v_1\ne v_2$ such that $v_1$ and $v_2$ are incident to $C_1$ and $C_2$, respectively~\cite{plane_gtwisted_socg}. The following lemma %
implies that $v_1$ and $v_2$ are compatible.

\begin{figure}[ht]
	\centering
	\includegraphics[scale=0.575,page=1]{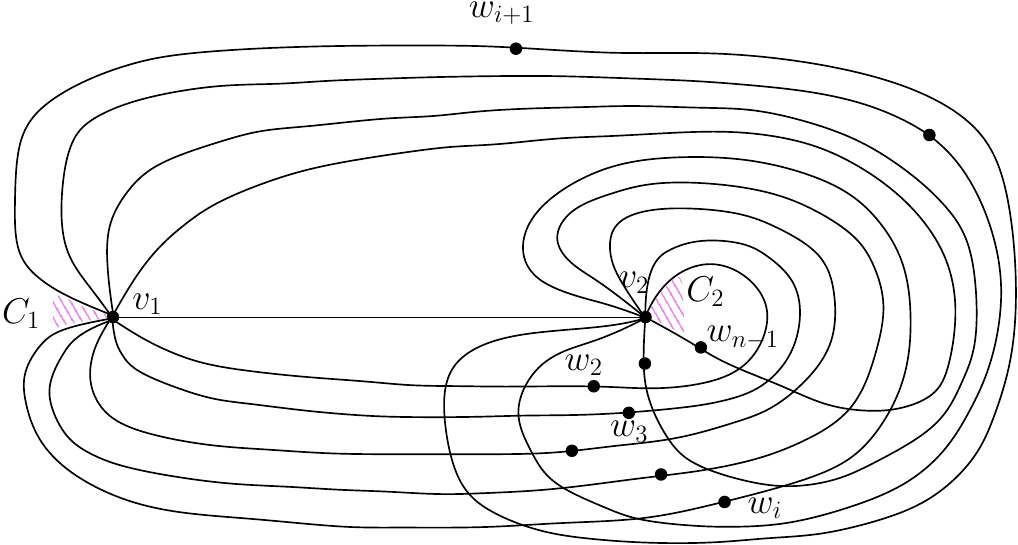}
	\caption{The rotations of $v_1$ and $v_2$.}
	\label{fig:rotations}
\end{figure}

\begin{figure}[thbp]
	\begin{minipage}[t]{.49\linewidth}
		\centering\includegraphics[scale=0.56,page=2]{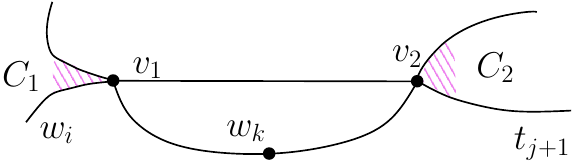}
	\end{minipage}
	\begin{minipage}[t]{.49\linewidth}
		\centering\includegraphics[scale=0.56,page=3]{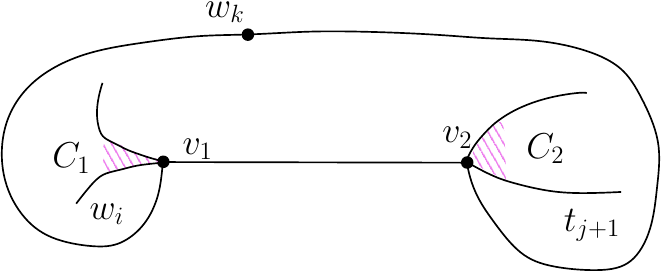}
	\end{minipage}
	\caption{$w_k$ cannot appear after $C_2$ around $v_2$.}\label{fig:lemmaw}
\end{figure}

\begin{lemma}\label{lem:compatible}
	Let $D$ be a realization of a generalized twisted rotation system of $K_n$ with~$n\geq 3$. Let $(C_1, C_2)$ be a pair of antipodal vi-cells of $D$, and let $v_1$ and $v_2$ be vertices of $D$ incident to $C_1$ and $C_2$, respectively, with $v_2\ne v_1$. Assume that $R_{v_1}=\{v_2=w_1, w_2, \ldots ,  w_{n-1}\}$ is the (clockwise) rotation of $v_1$, and that the edges $v_1w_i$ and $v_1w_{i+1}$ define part of the boundary of $C_1$, for some $i\in \{1,\ldots, n-1\}$. Then the following statements hold.
	\begin{enumerate}
		\item[(i)] If $i \in \{2, 3, \ldots, n-2\}$ then the (clockwise) rotation of $v_2$ is $\{v_1, w_i, w_{i-1}, \ldots , w_2,w_{n-1},$ $w_{n-2},$ $\ldots , w_{i+1}\}$ and the edges $v_2w_2$ and $v_2w_{n-1}$ define part of the boundary of $C_2$.
		\item[(ii)] If $i=1$ then the (clockwise) rotation of $v_2$ is $\{v_1, w_{n-1}, w_{n-2}, \ldots , w_{2}\}$ and the edges $v_2v_1$ and $v_2w_{n-1}$ define part of the boundary of $C_2$.
		\item[(iii)] If $i=n-1$ then the (clockwise) rotation of $v_2$ is $\{v_1, w_{n-1}, w_{n-2}, \ldots , w_{2}\}$ and the edges $v_2v_1$ and $v_1w_{2}$ define part of the boundary of $C_2$.
	\end{enumerate}
\end{lemma}

\begin{proof}[Proof Sketch]
We sketch the proof of \textsf{(i)}; the proofs of \textsf{(ii)} and \textsf{(iii)} are very similar.
Figure~\ref{fig:rotations} shows an example of the rotations of $v_1$ and $v_2$.
Assume that the clockwise rotation of~$v_2$ is $\{v_1=t_1, t_2, \ldots, t_{n-1}\}$ and that the edges $v_2t_j$ and $v_2t_{j+1}$, for some $j\in \{1,\ldots, n-1\}$, define part of the boundary of $C_2$.
	To place $w_k$ in the rotation of $v_1$ and $v_2$ with respect to $C_1$ and~$C_2$, we consider for $w_k$ with $2\le k \le i$ the triangle spanned by $v_1$, $v_2$ and $w_k$. As $C_1$ and $C_2$ are antipodal, they must lie on different sides of the triangle. This is not possible if $w_k$ corresponds to a vertex $t_{k'}$ with $j+1 \le k' \le n-1$ (see Figure~\ref{fig:lemmaw} for a depiction). Thus, if a vertex appears before (via analogous arguments after) $C_1$ around $v_1$ clockwise from $v_2$, then it also appears before (after) $C_2$ around $v_2$ clockwise from $v_1$.%

\begin{figure}[thbp]
	\begin{minipage}[t]{.49\linewidth}
		\centering\includegraphics[scale=0.55,page=4]{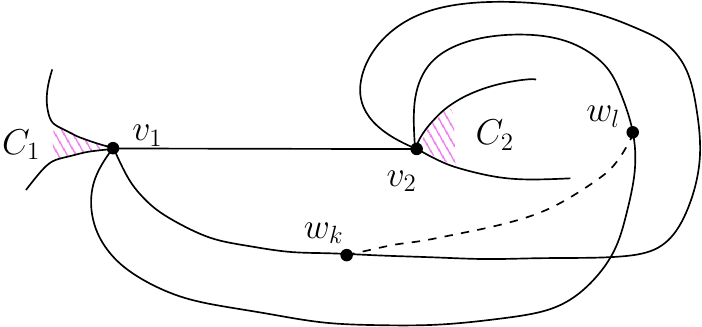}
	\end{minipage}
	\begin{minipage}[t]{.49\linewidth}
		\centering\includegraphics[scale=0.55,page=5]{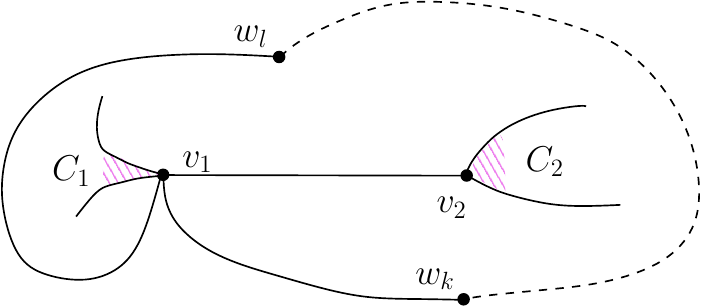}
	\end{minipage}
	\caption{$w_k$ cannot appear before $w_l$ around $v_2$.}\label{fig:lemma}
\end{figure}

For two vertices $w_k$ and $w_l$ with $2\le k < l \le i$, we show that $w_l$ appears before $w_k$ in the clockwise rotation of $v_2$ from $v_1$ by contradiction.
Consider the edges $v_1w_l$ and $v_2w_l$. If $w_k$ appears before $w_l$ in the clockwise rotation of $v_2$ from $v_1$, then the edges $v_1w_l$ and $v_2w_l$ emanate in different sides of the triangle $v_1v_2w_k$ from $v_1$ and $v_2$. From simplicity of the drawing and the fact that each realization of a generalized twisted rotation system of $K_4$ contains exactly one crossing~\cite{plane_gtwisted_journal}, we can determine how that edge $w_lw_k$ must behave (see Figure~\ref{fig:lemma} for a depiction). We then show that with this drawing, $C_1$ and $C_2$ must lie on the same side of the triangle~$ v_1w_kw_l$, which contradicts that they are antipodal cells.
\end{proof}

\begin{figure}[thbp]
	\begin{minipage}[t]{.49\linewidth}
		\centering\includegraphics[scale=0.58,page=6]{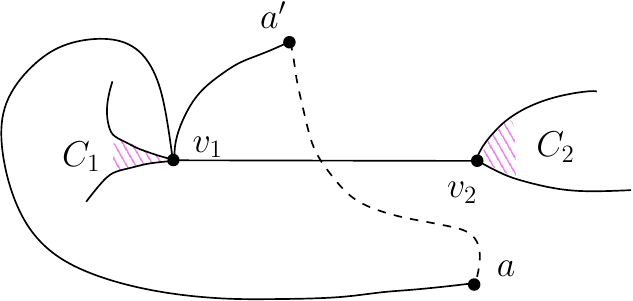}
	\end{minipage}
	\begin{minipage}[t]{.49\linewidth}
		\centering\includegraphics[scale=0.58,page=7]{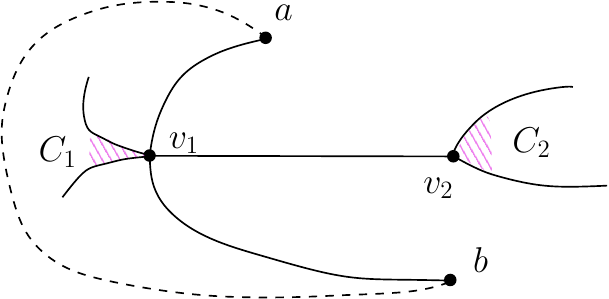}
	\end{minipage}
	\caption{Illustrating the proof of Theorem~\ref{the:bipartition}~\textsf{(i)} and \textsf{(iii)}.}\label{fig:iii}
\end{figure}

\begin{figure}[!htbp]
	\begin{minipage}[t]{.49\linewidth}
		\centering\includegraphics[scale=0.58,page=8]{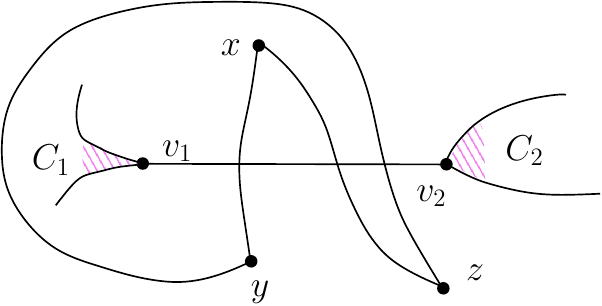}
	\end{minipage}
	\begin{minipage}[t]{.49\linewidth}
		\centering\includegraphics[scale=0.58,page=9]{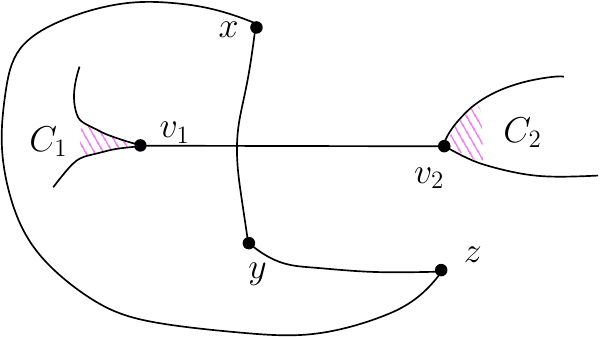}
	\end{minipage}
	\caption{$C_1$ and $C_2$ are on different sides of $\Delta xyz$. \textsf{Left:} $x$, $y$, and $z$
		belong to the same set of the bipartition. \textsf{Right:} $x$ and $y$ belong to one
		of the sets and $z$ to the other.}\label{fig:three}
\end{figure}

Using Lemma~\ref{lem:compatible} we can prove Theorem~\ref{the:bipartition}. 

\begin{proof}[Proof Sketch of Theorem~\ref{the:bipartition}]
	If $R$ is generalized twisted, then there exists a pair $(C_1, C_2)$ of antipodal vi-cells in $D$, and we show that a bipartition of the vertices satisfies \textsf{(i)}--\textsf{(v)}.
	Let $v_1$ be a vertex of $V$ incident to $C_1$ and let $v_2\ne v_1$ be a vertex of $V$ incident to $C_2$. Suppose that $\{v_2=w_1, w_2, \ldots ,  w_{n-1}\}$ is the (clockwise) rotation of $v_1$ such that $v_1w_i$ and $v_1w_{i+1}$ define part of the boundary of $C_1$, for some $i\in \{1,\ldots, n-1\}$.
	Let $A$ and $B$ be defined as follows: if $i \not \in \{1,n-1\}$, set $A:={w_{i+1}, ..., w_{n-1}}$, $B:={w_2, ..., w_i}$; if $i=1$, set $A:= \{v_1, w_{n-1}, w_{n-2}, \ldots , w_{2}\}$, $B := \emptyset$;
	if $i=n-1$, set $A:= \emptyset$, $B := \{v_1, w_{n-1}, w_{n-2}, \ldots , w_{2}\}$.
	We show that this bipartition satisfies \textsf{(i)}--\textsf{(v)}.

\begin{figure}[thbp]
	\begin{minipage}[t]{.49\linewidth}
		\centering\includegraphics[scale=0.58,page=10]{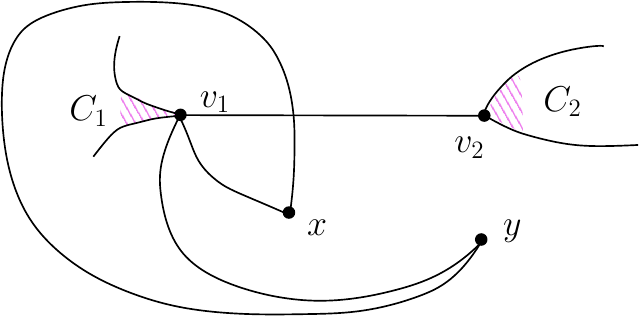}
	\end{minipage}
	\begin{minipage}[t]{.49\linewidth}
		\centering\includegraphics[scale=0.58,page=12]{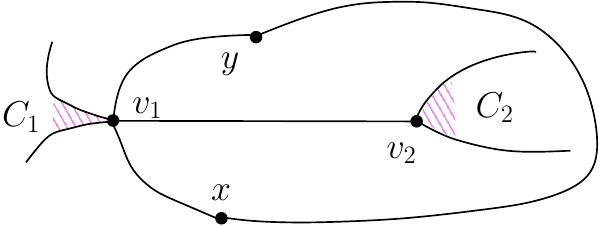}
	\end{minipage}
	\caption{$C_1$ and $C_2$ are on different sides of $\Delta v_1xy$.  \textsf{Left:} $x$ and $y$ belong to the
		same set of the bipartition.  \textsf{Right:} $x$ and $y$ belong to different sets of the bipartition.}\label{fig:two}
\end{figure}

\begin{figure}[thbp]
	\begin{minipage}[t]{.49\linewidth}
		\centering\includegraphics[scale=0.58,page=14]{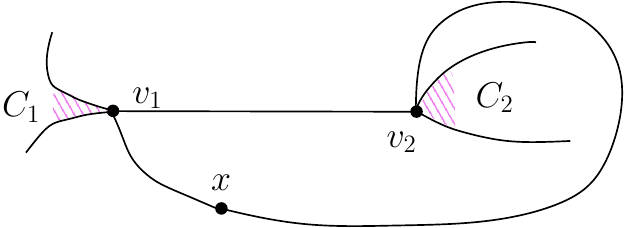}
	\end{minipage}
	\begin{minipage}[t]{.49\linewidth}
		\centering\includegraphics[scale=0.58,page=15]{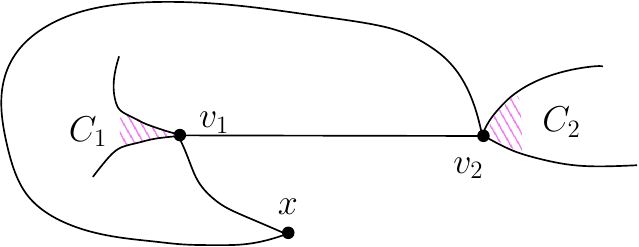}
	\end{minipage}
	\caption{$C_1$ and $C_2$ are on different sides of $\Delta v_1v_2x$.}\label{fig:one}
\end{figure}

To prove that \textsf{(i)} holds, we take two vertices $a$ and $a'$ in $A$. (If $A$ contains at most one vertex, \textsf{(i)} is vacuously true.) We use Lemma~\ref{lem:compatible} to obtain the rotation of $v_1$ and $v_2$ and information on the bounding edges of $C_1$ and $C_2$. We conclude that the edge $aa'$ crosses  $v_1v_2$ because otherwise $C_1$ and $C_2$ would be on the same side of the triangle $v_1aa'$ (see Figure~\ref{fig:iii}(left) for an example) contradicting that $C_1$ and $C_2$ are antipodal.
	Thus, \textsf{(i)} follows and, by analogous arguments for $B$, \textsf{(ii)}  follows.

	To prove that \textsf{(iii)} holds for $A$ and $B$, we take %
	vertices $a\in A$ and $b\in B$.
	By Lemma~\ref{lem:compatible}, $b$ is before $C_1$ and $a$ is after $C_1$ around $v_1$. Thus,  the edge $ab$ cannot cross $v_1v_2$ so that $C_1$ and $C_2$ are on different sides of $\Delta v_1ab$ (see Figure~\ref{fig:iii}(right) for an example). Hence, \textsf{(iii)} follows.
	Finally, \textsf{(iv)}--\textsf{(v)} follow directly from the definitions of $A$ and $B$.

	To prove the other direction, we assume that $R$ and $D$ are a rotation system and drawing fulfilling \textsf{(i)}--\textsf{(v)}, and show that $R$ is generalized twisted.
	Let $\{v_2, b_1, \ldots, b_{n_1},$ $a_1, \ldots , a_{n_2}\}$ be the (clockwise) rotation of $v_1$ beginning at $v_2$, and $\{v_1, b'_1, \ldots, b'_{n_1},$ $a'_1, \ldots , a'_{n_2}\}$ be the (clockwise) rotation of $v_2$ beginning at $v_1$. Let $C_1$ be the vi-cell of $D$ defined by the edges $v_1b_{n_1}$ and $v_1a_1$, and $C_2$ be the vi-cell of $D$ defined by the edges $v_2b'_{n_1}$ and $v_2a'_1$. We show that $C_1$ and $C_2$ are antipodal and thus $R$ is generalized twisted by showing for all possible triangles in $D$ that $C_1$ and $C_2$ are on different sides of that triangle.
	We analyze the triangles and show the following.
	If $v_1$ and $v_2$ are not incident to $\Delta$, the number of crossings give that $v_1$ and $v_2$ -- and thus $C_1$ and $C_2$ -- are on different sides of $\Delta$ (see Figure~\ref{fig:three}). If $v_1$ or $v_2$ are incident to $\Delta$, %
	we use $C_1$ and $C_2$ being on different ends of $v_1v_2$ to show that $C_1$ and $C_2$ are on different sides of  $\Delta$ (see Figures~\ref{fig:two} and~\ref{fig:one}).
\end{proof}

\section{Efficient algorithms}\label{sec:algo}
In this section, we turn to the algorithmic aspect and prove Theorem~\ref{thm:algo}.

To check whether an abstract rotation system of $K_n$ is generalized twisted, by Theorem~\ref{thm:T5}, only the subrotation systems induced by five vertices need 
checking.
This can be done in $O(n^5)$ time. (The only exception where subdrawings are not sufficient are abstract rotation systems of~$K_6$, but for $n=6$ all three generalized twisted rotation systems are known.)
Alternatively, one can first check in $O(n^5)$ time whether the rotation system is realizable~\cite{all_small_drawings,realizability}. If it is not, it cannot be generalized twisted. If it is,
we can use the following faster algorithm for realizable rotation systems of $K_n$, which we developed using Theorem~\ref{the:bipartition}.
The algorithm runs in $O(n^2)$ time and in two steps, which we describe in the remainder of this section.

\medskip

\noindent \textsl{\textbf{Step 1.}}
In the first step, the empty star triangles at an arbitrary vertex $x$ are computed.

\begin{lemma}\label{lem:algo:step1}
	Given a realizable rotation system $R$ of $K_n$ and a vertex $x$ of~$R$, the set of all empty star triangles at $x$ is the same in each realization of $R$ and can be computed from $R$ in $O(n^2)$ time.
\end{lemma}

\begin{proof}[Proof Sketch]
	By Lemma~\ref{lem:gen_startriangles}, a star triangle $xyz$ at $x$ is empty if and only if $y$ and $z$ are consecutive in the rotation of~$x$. Hence, there are at most $n-1$ empty star triangles at $x$. Whether a triangle $xyz$ is a star triangle at~$x$ is determined by whether there exists a vertex $w$ such that $xw$ crosses~$yz$. Since deciding for two edges if they cross each other can be done in constant time via the rotation system, deciding if a star triangle is empty requires $O(n)$ %
	time. Thus, computing all empty star triangles at $x$ can be done in $O(n^2)$ time.
\end{proof}

\medskip

\noindent \textsl{\textbf{Step 2.}}
In the second step, the previously computed empty star triangles at $x$ and Theorem~\ref{the:bipartition} are used to decide whether $R$ is generalized twisted. %

\begin{lemma}\label{lem:algo:step2}
	Given a realizable rotation system $R$ of $K_n$ and the set of empty star triangles of a %
	vertex $x$ of $R$, it can be determined in $O(n^2)$ time whether  $R$ is generalized twisted.
\end{lemma}

\begin{proof}[Proof Sketch]
	By Lemma~\ref{lem:compatible}, every drawing with generalized twisted rotation system has a pair $(v_1,v_2)$ of compatible vertices. %
If $(C_1, C_2)$ is a pair of antipodal vi-cells, then $v_1$ and~$v_2$ lie on the boundary of $C_1$ and $C_2$ (or vice versa).
We look for compatible vertices using empty triangles.
Let $x$ be the vertex with given empty star triangles. If there are not exactly two empty star triangles at $x$, then $R$ is not generalized twisted by Lemma~\ref{lem:gtwisted_startriangles}. Otherwise, we denote by $\Delta = xyz$ and $\Delta' = xy'z'$ the two empty star triangles at~$x$.
(Possibly, $y$ or $z$ coincides with $y'$ or $z'$.)
Using %
Lemma~\ref{lem:gtwisted_startriangles}, we can show that if there is a compatible pair of vertices it must be in $S=\{(y,y'), (y,z'), (y,x), (z,y'), (z,z'),$ $(z,x), (x,y'), (x,z')\}$.
Checking compatibility of any pair in $S$  by comparing their rotation systems takes linear time.

	For every such potentially compatible pair $(v_1,v_2)$, we check properties \textsf{(i)}--\textsf{(v)} of Theorem~\ref{the:bipartition}.
Let $\{v=w_1, w_2, \ldots ,  w_{n-1}\}$ be the rotation of $v_1$. We define $A$ and $B$ as follows. If the rotation of $v_2$ is $\{v_1, w_{n-1}, \ldots , w_{2}\}$, then $B = \{w_2, \ldots , w_{n-1}\}$ and $A=\emptyset$ (or vice versa), %
and if the rotation of $v_2$ is $\{v_1, w_i, \ldots , w_2,w_{n-1}, \ldots , w_{i+1}\}$, for some $i$ different from $1$ and $n-1$, then $B=\{w_2, \ldots , w_i\}$ and $A=\{w_{i+1}, \ldots , w_{n-1}\}$.
Then $A$ and $B$ clearly satisfy \textsf{(iv)} and \textsf{(v)} of Theorem~\ref{the:bipartition}.
Checking if \textsf{(i)}, \textsf{(ii)}, and \textsf{(iii)} of Theorem~\ref{the:bipartition} are satisfied requires $O(n^2)$ time.

Therefore, since $|S|$ is a constant, deciding if there is a pair $(v_1,v_2)$ in $S$ such that $v_1$ and $v_2$ are compatible and the properties \textsf{(i)}--\textsf{(v)} of Theorem~\ref{the:bipartition} are satisfied for this pair can be done in $O(n^2)$ time.
\end{proof}

\section{Conclusion and open problems}\label{sec:conclusion}

We presented two new characterizations for generalized twisted drawings based solely on their rotation systems. From these, we derived an $O(n^5)$-time algorithm that decides whether an abstract rotation system of~$K_n$ is generalized twisted and an $O(n^2)$-time algorithm to decide whether a realizable rotation system is generalized twisted. The obvious open question is whether the decision can also be done faster for abstract rotation systems. The quadratic time algorithm we give does not directly apply to them. There is  a non-realizable rotation system of $K_5$ for which the output of the algorithm depends on the choice of vertex $x$ and might be 'generalized twisted' or 'not generalized twisted'.

Generalized twisted drawings have %
been useful before, for example to find large plane substructures in simple drawings~\cite{plane_gtwisted_journal,triangle_gtwisted_journal}.
Overcoming their rather geometrical definition, our new combinatorial characterizations might ease work with generalized twisted drawings and foster further results with this flavor.
For example, Pach, Solymosi, and T{\'o}th~\cite{pachetal} showed that every simple drawing of a complete graph on $n$ vertices contains $O(\log^{\frac{1}{8}}n)$ vertices	inducing a complete subdrawing that is either convex or twisted.
Recently, Suk and Zeng~\cite{unavoidableNewCounference,unavoidableNewJournal} improved this result to $(\log n)^{\frac{1}{4} - o(1)}$ vertices.
As both convex and twisted drawings contain large plane paths and cycles, this provides lower bounds on the size of plane paths and cycles that every simple drawing contains. %
It would be interesting to generalize this to show that every simple drawing contains a large generalized twisted or generalized convex subdrawing. Generalized convex drawings are -- like generalized twisted drawings -- a family of drawings of graphs that are well investigated~\cite{gconvex_char,gconvex_cycles,convex_holes}. Similarly to generalized twisted drawings, they admit characterizations via small constant sized subrotation systems~\cite{gconvex_char}.
Thus, a result guaranteeing the existence of a large generalized convex or generalized twisted drawing would help to extend our knowledge on simple drawings and, for example, to improve the lower bounds on the size of plane paths and cycles.
We expect that this and several other questions on simple drawings can be tackled using the characterizations of this paper -- independently or in combination. %
Especially, since the two given characterizations are very different from each other, with Theorem~\ref{thm:T5} being on all small subrotation systems and Theorem~\ref{the:bipartition} being on two specific points and their rotation in the whole rotation system, they might foster the application of different other tools to obtain new results.

\newpage
\bibliography{twisted_rotations}
\newpage

\appendix
\section*{Appendix}

\section{Missing proofs of Section~\ref{sec:pre}}

\begin{proof}[Proof of Lemma~\ref{lem:only_one_maxcross}]
	Consider first the case $n=5$. There are (up to strong isomorphism) only five simple drawings of $K_5$~\cite{ahprsv-etgdc-15}. %
Among them, the only drawing containing a maximum crossing edge is the corresponding to the {\gtwisted} rotation system %
$T_5$ of~$K_5$ (see Figure~\ref{sfig:lemmaw2}), which contains exactly one maximum crossing edge. Thus, there is no drawing of $K_5$ that contains more than one maximum crossing edge.

Consider now the case $n>5$. Assume, for a contradiction, that there is a drawing $D$ with two maximum crossing edges, say $uv$ and $u'v'$ (some of the vertices can coincide). Take an induced subdrawing of five vertices of $D$ containing~$u,v,u',v'$. Since $uv$ crosses all edges of $D$ not in $S(u) \cup S(v)$ and $u'v'$ crosses all edges of $D$ not in $S(u') \cup S(v')$, it follows that both edges are maximum crossing also in this drawing of $K_5$. However, there is only one maximum crossing edge in any drawing of $K_5$, a contradiction. We conclude that for $n>5$, there is also no drawing of $K_n$ that contains more than one maximum crossing edge.
\end{proof}

\begin{proof}[Proof of Lemma~\ref{lem:maxcross_then_gtwisted}]

	By definition, $C_1, C'_1, C_2$ and $C'_2$ are vi-cells. Thus, we only need to show by Theorem~\ref{thm:antipodal_gtwisted}  %
that $(C_1, C_2)$ and $(C'_1, C'_2)$ are pairs of antipodal cells. We only prove that $(C_1, C_2)$ is a pair of antipodal cells, as the proof for $(C'_1, C'_2)$ is totally analogous. %

Consider a triangle $xyz$ with $\{x,y,z\}\bigcap \{v_1,v_2\} = \emptyset $. Since all edges on the boundary of $xyz$ cross $v_1v_2$, the vertices $v_1$ and $v_2$ have to lie on different sides of the triangle. Then $C_1$ lies on one side of the triangle, while $C_2$ lies on the other side. For a triangle $v_1yz$ with $v_2 \notin \{y,z\}$, the edge $v_1v_2$ crosses exactly one edge (the edge $yz$) on the boundary of $v_1yz$. Thus, the edge $v_1v_2$ has to emerge from $v_1$ at the other side of the triangle than $v_2$ lies and hence, the cell $C_1$ lies on the other side of the triangle than $v_2$ and therefore $C_2$. Analogously, $C_1$ and $C_2$ lie on different sides of any triangle $v_2yz$ with $v_1 \notin \{y,z\}$. Finally, as by definition $C_1$ and $C_2$ lie on different sides of the edge $v_1v_2$, necessarily for any triangle $v_1v_2x$, $C_1$ lies on the other side of the triangle than $C_2$. In conclusion, $C_1$ and $C_2$ are antipodal and $D$ is {\gtwisted}.
\end{proof}

\section{Missing proofs of Section~\ref{sec:5}}\label{app:sec:5}

\begin{proof}[Proof of Lemma~\ref{lem:exactly_two}]
Let $v$ be a vertex of $D$. By Lemma~\ref{lem:gen_startriangles}, there are at least two empty star triangles at $v$. We show that there are exactly two empty star triangles at $v$. Assume for a contradiction that there are at least three empty star triangles at~$v$. The subdrawing $D'$ induced by the vertices of the triangles (at least five vertices) has a {\gtwisted} rotation system by hypothesis. %
As all empty star triangles of $D$ stay empty star triangles when vertices are removed (unless a vertex of the triangle is removed), then $v$ is a vertex in a drawing $D'$ whose rotation system is {\gtwisted} such that there are at least three empty star triangles at $v$. This contradicts Lemma~\ref{lem:gtwisted_startriangles}. Therefore, there are exactly two empty star triangles at any vertex of $D$.
\end{proof}

\begin{proof}[Proof of Lemma~\ref{lem:adjacent_then_maxcross}]
Let the empty star triangles at $v_1$ be $\Delta=v_1v_2v_3$ and $\Delta' = v_1v_2v'_3$.
As shown in~\cite{plane_gtwisted_journal}, the realization of a {\gtwisted} rotation system of~$K_4$ is (up to strong isomorphism) unique~\cite{plane_gtwisted_socg} and necessarily has a crossing; see Figure~\ref{fig:Case1_1} for a depiction of the two empty star triangles where the extension to a $K_4$ is drawn dashed.
By definition of star triangle, the edge $v_1v'_3$ cannot cross $v_2v_3$ and the edge $v_1v_3$ cannot cross $v_2v'_3$. Thus, the only pair of edges that is allowed to cross is $v_3v'_3$ and~$v_1v_2$. By construction, this crossing has to lie on the empty side of the quadrilateral $Q=v_1v_3v_2v'_3$.

\begin{figure}[!tb]
	\centering
		\subfloat[]{%
				\includegraphics[scale=0.76,page=1]{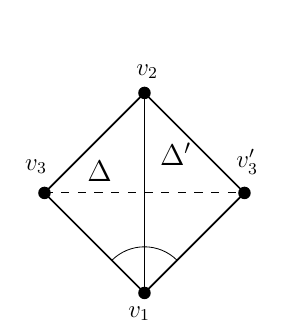}
				\label{fig:Case1_1}
			}~~~
			\subfloat[]{
				\includegraphics[scale=0.76,page=1]{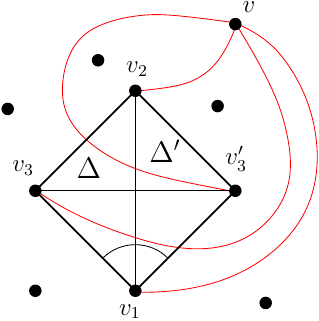}
				\label{fig:Case1_2}
			}~~~
			\subfloat[]{
				\includegraphics[scale=0.76,page=1]{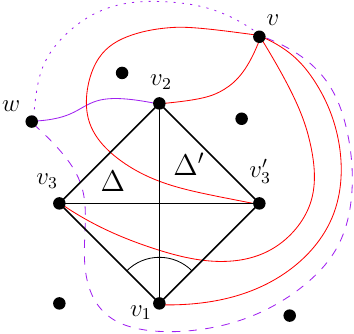}
				\label{fig:Case1_3}
			}
			\caption{Illustrating the proof of Lemma~\ref{lem:adjacent_then_maxcross}. (a) Two adjacent empty star triangles $\Delta$ and $\Delta'$ at $v_1$. (b) The {\gtwisted} drawing $D_0$. (c) The edge $vw$ must cross $v_1v_2$.}
			\label{fig:adjacent_triangles}
		\end{figure}

Let $v$ be a vertex not of $Q$ and consider the drawing $D_0$ induced by $v$ and the vertices of~$Q$. Since the rotation system of $D$ is {\gtwisted}, the rotation system of $D_0$ is also {\gtwisted}. As the only {\gtwisted} rotation system on 5 vertices is $T_5$, then $D_0$ is strongly isomorphic to the drawing in Figure~\ref{fig:triangle_vi}, where $v_1$ is either $e$ or $c$ because they are the only vertices with the two star triangles being adjacent. Thus, the only possibilities for $D_0$ are the drawing depicted in Figure~\ref{fig:Case1_2} and its mirror image (with $vv_3$ crossing~$v_2v'_3$, $vv'_3$ crossing $v_1v_3$, and both of $vv_3$ and $vv'_3$ crossing $v_1v_2$).
In particular, $vv_3$ and $vv'_3$ must cross $v_1v_2$, regardless of the vertex $v$.
On the other hand, in the rotation of $v_2$, the vertices $v'_3$ and $v_1$ are consecutive, as are $v_1$ and $v_3$.
Since this holds for any $v$, $\Delta$ and $\Delta'$ are also star triangles at $v_2$, which is Property~\ref{lem:adjacent_then_maxcross_ii} of the lemma.
Further, in the rotation of $v_3$ and the rotation of $v'_3$, the edges to $v_1$ and $v_2$ are consecutive in such a way that all edges emanate from $v_3$ and $v'_3$ in the empty side of the triangles $\Delta$ and $\Delta'$, respectively.
Since this again holds for any $v$, this completes the proof of Property~\ref{lem:adjacent_then_maxcross_iii} of the lemma.

For Property~\ref{lem:adjacent_then_maxcross_i}, it remains to show that
also all edges not incident to vertices in $Q$ cross $v_1v_2$.
Assume, for a contradiction, that there are such vertices $v$ and $w$ such that the edge $vw$ does not cross the edge $v_1v_2$.
Since $Q$ consists of the two empty triangles $\Delta$ and $\Delta'$, both $v$ and $w$ lie outside of $Q$.
Note that as $\Delta$ and $\Delta'$ are empty star triangles at $v_2$, an edge $v_2z$, with $z\notin \{v_1, v_3, v'_3\}$, cannot cross $Q$ and must be outside $Q$. We fist show that $vw$ has to cross $Q$.
If $vw$ does not cross $Q$ at all, then both of $\Delta$ and $\Delta'$ lie on one side of $v_2vw$ (see Figure~\ref{fig:Case1_3}, where $vw$ is depicted dotted). Hence the triangles $\Delta$, $\Delta'$, and $v_2vw$ have pairwise disjoint sides, a contradiction to $D$ being {\gtwisted} by Lemma~\ref{lem:threetriangles}.
So assume that $vw$ crosses $Q$ but does not cross $v_1v_2$.
Then it has to cross either bot of $v_3v_1$ and $v_3v_2$, or both of $v'_3v_1$ and $v'_3v_2$.
However, since $v_1$ and $v_2$ are consecutive around $v_3$ by Property~\ref{lem:adjacent_then_maxcross_iii}, the edge $vw$ cannot cross both of $v_3v_1$ and $v_3v_2$ without crossing also $v_3v$ in between, which it cannot by simplicity (see Figure~\ref{fig:Case1_3}, where $vw$ is depicted dashed).
Analogously, $vw$ cannot cross both of $v'_3v_1$ and $v'_3v_2$. It follows that $vw$ must cross $v_1v_2$.
In conclusion, any edge that is not incident to $v_1$ or $v_2$
crosses $v_1v_2$. Thus, $v_1v_2$ is maximum crossing, which is Property~\ref{lem:adjacent_then_maxcross_i} of the lemma and concludes the proof.
\end{proof}

\begin{figure}[!tb]
	\centering
	\begin{subfigure}{.45\textwidth}
		\includegraphics[scale=0.58,page=2]{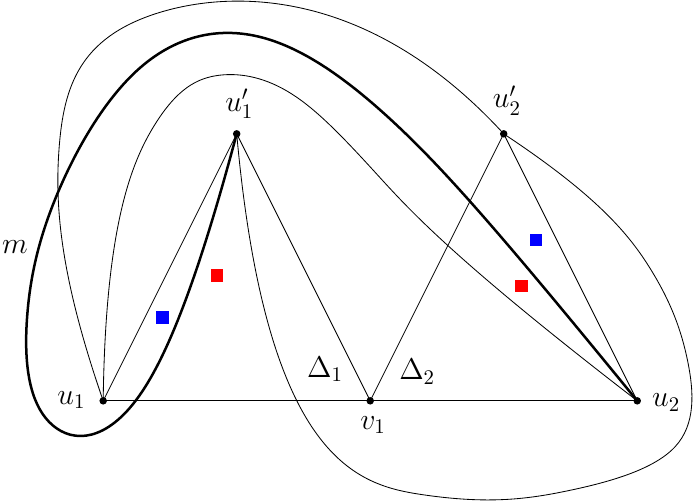}
	\end{subfigure}
	\hfill
	\begin{subfigure}{.45\textwidth}
		\includegraphics[scale=0.58,page=3]{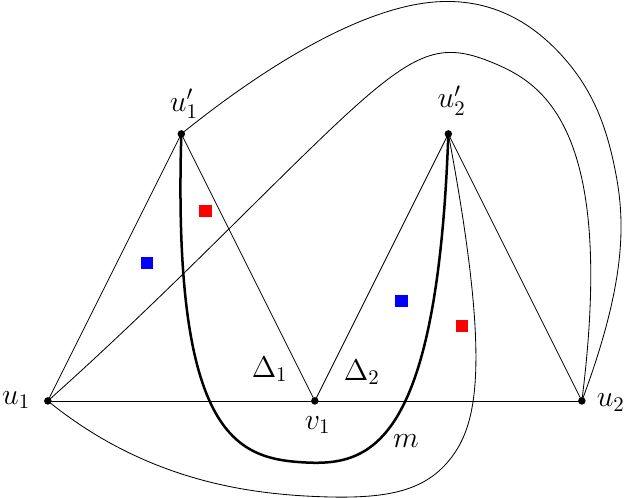}
	\end{subfigure}
	\caption{The (up to strong isomorphism and mirroring) only two possibilities to draw two non-adjacent empty star triangles at $v_1$ such that $v_1$ is in the center. The (bold) edge $m$ is the maximum crossing edge of this drawing of $K_5$. (Left: $m=u'_1u_2$ crosses all edges not adjacent to it. Right: $m=u'_1u'_2$ cross all edges adjacent to it.)}
	\label{appfig:case2}
\end{figure}

\subparagraph*{Remarks on the similarity of $\Delta_1$ and $\Delta_2$.}
Figure~\ref{appfig:case2} shows again shows the two choices of $v_1$ in such a way that $v_1$ is in the center of the drawing. The labeling of $\Delta_1$ and $\Delta_2$ is chosen without loss of generality (in particular, $\Delta_2$ could be the left triangle and everything would behave the same). Figure~\ref{appfig:case2}(left) shows the case where $v_1$ is $b$ and Figure~\ref{appfig:case2}(right) the case where $v_1$ is $a$ or~$d$. During most of the proof, we will argue via the triangles $\Delta_1$ and $\Delta_2$ separately, but simultaneously, because we will consider only one triangle, but need that the other triangles has the same properties. In the case that $v_1$ is not incident to any via-cells, as in Figure~\ref{appfig:case2}(right), the triangles $\Delta_1$ and $\Delta_2$ behave analogously. In case that $v_1$ is incident to a via-cell, it is incident to exactly one via-cell in one of the triangles. In the labeling we chose without loss of generality in Figure~\ref{appfig:case2}(right), $v_1$ is adjacent to a via-cell in~$\Delta_2$. In that case, $\Delta_1$ and $\Delta_1$ do not behave completely analogously, but all needed properties hold by analogous proofs for $\Delta_1$ and $\Delta_2$.

Before proving Lemma~\ref{lem:two_pairs_adjacent} and Lemma~\ref{lem:included_cells}, we state the following results, of which three we cite from published work~\cite{triangle_gtwisted_journal}, one is a direct corollary, another is an observation and two we state and prove here.

\begin{lemma}[\cite{triangle_gtwisted_journal}]\label{lem:gtwisted_triangles_special}\label{lem:gtwisted_triangles}	
	Let $D$ be a generalized twisted drawing of $K_n$, let $(C_1,C_2)$ be a pair of antipodal vi-cells, and let $v_1$ a vertex on the boundary of $C_1$. Then the triangle $\Delta_{v_1}$ defined by the two edges incident to $v_1$ and $C_1$ is an empty star triangle at $v_1$. The triangle $\Delta$ is also an empty star triangle at another vertex, say $x$. For the vertex of $\Delta_{v_1}$ that is different from $x$ and $v_1$ it holds that any edge incident to that vertex crosses $v_1x$. Further, let $ab$ be an edge different from $v_1x$ and $v_1y$ that is on the boundary of $C_1$. Then the triangle $abv_1$ is an empty star triangle at $a$ and $b$ with $C_1$ on its empty side.
\end{lemma}

\begin{lemma}[\cite{triangle_gtwisted_journal}]\label{lem:gtwisted_triangles_inside}	
	Let $D$ be a generalized twisted drawing of $K_n$, let $C$ be a via-cell of $D$, let $v$ be a vertex incident to $C$, let $\Delta$ be an empty triangle in $D$ that has $C$ on its empty side. Then the following holds:
	\begin{enumerate}
		\item \label{emptyv} The vertex $v$ is a vertex of $\Delta$, that is, $\Delta=xyv$ for some $x,y$.
		\item \label{empty1or2} %
		The triangle $\Delta=xyv$ is an empty star triangle at $x$ or $y$ or both.
		\item \label{emptyno3}
		If $\Delta=xyz$ is a star triangle at two vertices, say $x$ and $y$,
		then all edges from the third vertex~$z$, except $xz$ and $yz$, emanate from $z$ to the empty side of $\Delta$ and cross $xy$.
	\end{enumerate}
\end{lemma}

\begin{lemma}[\!\cite{triangle_gtwisted_journal}]\label{lem:threetriangles}
	Let $D$ be a simple drawing of $K_n$ with $n\ge 4$ that is strongly isomorphic to a  {\gtwisted} drawing. Then $D$ does not contain three interior-disjoint triangles.
\end{lemma}

The following corollary is a direct consequence of Theorem~\ref{thm:antipodal_gtwisted}.

\begin{corollary}%
	\label{cor:antipodal_gtwisted}
	A simple drawing $D$ of $K_n$ with $n\ge 3$ contains a pair of antipodal vi-cells if and only the rotation system of $D$ is {\gtwisted}.
\end{corollary}

\begin{figure}[!tb]
			\centering
				\subfloat[]{%
						\includegraphics[scale=0.9,page=1]{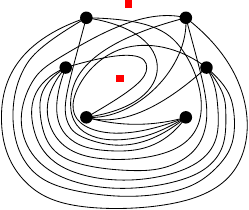}
						\label{fig:K6_1}
					}~~~~~~~~
					\subfloat[]{
						\includegraphics[scale=0.9,page=1]{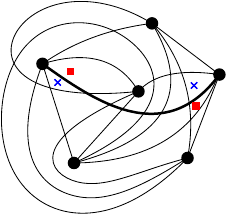}
						\label{fig:K6_2}
					}~~~~~~~~
					\subfloat[]{
						\includegraphics[scale=0.9,page=1]{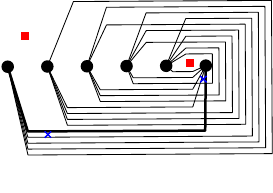}
						\label{fig:K6_3}
					}
					\caption{A realization for each {\gtwisted} rotation system of $K_6$, with marked pairs of antipodal vi-cells.}
					\label{appfig:gtwistedk6}
		\end{figure}

There is one unique generalized twisted rotation system of $K_5$ and three generalized twisted rotation systems of~$K_6$~\cite{plane_gtwisted_journal} (see Figure~\ref{appfig:gtwistedk6} for depictions of realizations as in~\cite{triangle_gtwisted_journal}). When analyzing these rotation systems and their realizations, we can observe the following.

\begin{observation}\label{obs:K6}
	Let $D$ be a generalized twisted drawing of~$K_5$ or~$K_6$. Then $D$ contains at most two pairs of antipodal vi-cells. If $D$ contains two pairs of antipodal vi-cells, say $(C_1,C_2)$ and $(C'_1,C'_2)$, then there are vertices $v_1$, $v_2$ such that: (1)~$v_1$ lies on the boundary of~$C_1$ and either on the boundary of $C'_1$ or the boundary of~$C'_2$. (2)~The vertex $v_2$ lies on the boundary of $C_2$ and on the boundary of a cell in $(C'_1,C'_2)$ that does not have $v_1$ on its boundary. %
	(3)~The edge $v_1v_2$ is maximum crossing.
\end{observation}

\begin{lemma}\label{lem:containment}
	Let $D$ be a simple drawing of $K_n$ with $n\ge 6$ whose rotation system is {\gtwisted}, and $(C_1,C_2)$ be a pair of antipodal vi-cells. Then for any vertex $v$ not on the boundary of $C_1$ or $C_2$,
	$(C_1,C_2)$ is contained in a pair $(\bar{C}_1, \bar{C}_2)$ of antipodal vi-cells of~$D\!\setminus\! \{v\}$.
\end{lemma}

\begin{proof}

	We first observe the following about simple drawings. Let $D$ be a simple drawing and let $V$ be a set of vertices in $D$. Then each cell $C$ of $D$ is contained in a cell in $D\!\setminus\! V$. Let $v \notin V$ be a vertex on the boundary of a cell $C$ in $D$ and let $\bar{C}$ be the cell in $D\! \setminus\! V$ that contains $C$. Then $v$ lies on the boundary of $\bar{C}$ in $D\! \setminus\! V$.
	
	Now, let $\bar{C}_1$ and $\bar{C}_2$ be the cells in $D\!\setminus\! \{v\}$ that contain $C_1$ and $C_2$, respectively. Since $C_1$ and $C_2$ are vi-cells and $v$ is not on the boundary of either $C_1$ or $C_2$, the cells $\bar{C}_1$ and $\bar{C}_2$ are vi-cells in $D\!\setminus\! \{v\}$. Besides, they are antipodal, as any triangle in $D$ separating $C_1$ and $C_2$, necessarily has to separate $\bar{C}_1$ and $\bar{C}_2$ in $D\!\setminus\! \{v\}$. Thus, $(\bar{C}_1, \bar{C}_2)$ is a pair of antipodal vi-cells in $D\!\setminus\! \{v\}$.
\end{proof}

\begin{lemma}\label{lem:two_non_adjacent}
	Let $D$ be a simple drawing of $K_n$ with $n\ge 6$ whose rotation system is {\gtwisted}. Suppose there is a vertex $v$ such that the two empty star triangles $vx_1x_2$ and $vy_1y_2$ at $v$ are not adjacent. Let $(\bar{C}_1,\bar{C}_2)$ and $(\bar{C}'_1,\bar{C}'_2)$ be the two pairs of antipodal vi-cells of the subdrawing $D_0$ induced by $v, x_1, x_2, y_1$ and $y_2$. Then for any pair $(C,C')$ of antipodal vi-cells of $D$, $(C,C')$ is contained in either $(\bar{C}_1,\bar{C}_2)$ or $(\bar{C}'_1,\bar{C}'_2)$.
\end{lemma}

\begin{proof}
	Since the rotation system of $D$ is {\gtwisted}, by Lemma~\ref{lem:gtwisted_startriangles} there are exactly two empty star triangles at any vertex of $D$, and if $(C,C')$ is a pair of antipodal vi-cells of $D$, then one of the two cells is in one of the empty star triangles and the other cell in the other empty star triangle. Thus, one of $C$ and $C'$ is in the triangle $vx_1x_2$ and the other one in the triangle $vy_1y_2$. In particular, this implies that any vertex $v$ different from $v_1, x_1, x_2, y_1$ and $y_2$ cannot belong to the boundary of either $C$ or $C'$ because $vx_1x_2$ and $vy_1y_2$ are empty. Thus, by successively removing the vertices of $D$ not in $D_0$ and repeatedly applying Lemma~\ref{lem:containment}, necessarily $(C,C')$ is contained in either $(\bar{C}_1,\bar{C}_2)$ or $(\bar{C}'_1,\bar{C}'_2)$.
\end{proof}

\begin{proof}[Proof of Lemma~\ref{lem:two_pairs_adjacent}]
Note that by Corollary~\ref{cor:antipodal_gtwisted}, the rotation system of $D$ is {\gtwisted}, so the rotation system of any subdrawing induced by any set of vertices is also {\gtwisted}.

\begin{figure}[!htb]
	\centering
		\subfloat[$v_1=a$.]{%
				\includegraphics[scale=0.86,page=1]{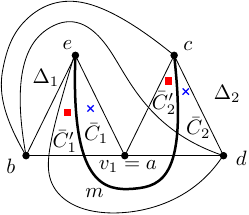}
				\label{fig:Case2v_1}
			}~~~~~~~~~~~
			\subfloat[$v_1=d$.]{
				\includegraphics[scale=0.86,page=1]{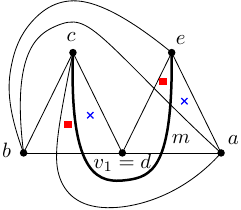}
				\label{fig:Case2v_2}
			}~~~~~~~~~~~
			\subfloat[$v_1=b$.]{
				\includegraphics[scale=0.86,page=1]{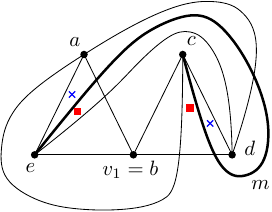}
				\label{fig:Case2v_3}
			}
			\caption{Realizations of $T_5$ such that the vertex $v_1$ which is incident to two non-adjacent empty star triangles at $v_1$ is drawn in the center. The label corresponds to the labels of Figure~\ref{sfig:lemmaw2}.}
			\label{appfig:v_center}
		\end{figure}
		
		We start proving Property~\ref{slem:p1}. Recall that $C_1, C_2, C'_1$ and $C'_2$ must be different. Let $v_1$ be one of the vertices on the boundary of $C_1$, $v_2$ on $C_2$, $v'_1$ on $C'_1$ and $v'_2$ on $C'_2$. For each of these vertices, there are exactly two empty star triangles by Lemma~\ref{lem:gtwisted_startriangles}. If any such pair of empty star triangles shares an edge, we are done. So assume for a contradiction that there is no vertex such that its two empty star triangles share an edge.
		
		Let $\Delta_1$ and $\Delta_2$ be the two empty star triangles at $v_1$. Let $D_0$ be the {\gtwisted} subdrawing induced by the five vertices in $\Delta_1$ and $\Delta_2$. Since $\Delta_1$ and $\Delta_2$ are non-adjacent empty star triangles in $D$, they also are non-adjacent empty star triangles in $D_0$. There is only a {\gtwisted} rotation system $T_5$ of $K_5$, so $D_0$ is strongly isomorphic to the subdrawing in Figure~\ref{sfig:lemmaw2}, where angles leading to an empty star triangle are marked by circular arcs. Two of the vertices, $c$ and $e$, have adjacent empty star triangles, so $v_1$ has to be one of the other vertices. Displays of the three possible cases for $v_1$ can be seen in Figure~\ref{appfig:v_center}, where $v_1$ is drawn in the center. Let $(\bar{C}_1,\bar{C}_2)$ and~$(\bar{C}'_1,\bar{C}'_2)$ be the two pairs of antipodal vi-cells in $D_0$. By Lemma~\ref{lem:two_non_adjacent}, $(C_1,C_2)$ and~$(C'_1,C'_2)$ must be contained in either $(\bar{C}_1,\bar{C}_2)$ or~$(\bar{C}'_1,\bar{C}'_2)$. As $b$ is not on the boundary of any via-cell in $D_0$, then $v_1$ cannot be $b$. Thus, $v_1$ must be either $a$ or $d$.  Since the properties of the drawing with respect to these two vertices, $a$ and $d$, are analogous (see Figures~\ref{fig:Case2v_1} and~\ref{fig:Case2v_2}),   %
		we may assume without loss of generality that $v_1=a$, as in Figure~\ref{fig:Case2v_1}.
		We may also assume that $\Delta_1$ is $v_1eb$, $\Delta_2$ is $v_1cd$, and $(\bar{C}_1, \bar{C}_2)$ (cells marked with blue crosses in Figure~\ref{fig:Case2v_1}) is the pair of antipodal vi-cells of $D_0$ containing $(C_1, C_2)$ such that $\bar{C}_1$ is in $\Delta_1$, $\bar{C}_2$ is in $\Delta_2$ and $v_1$ is on the boundary of $C_1$ and $\bar{C}_1$. Recall that by Lemma~\ref{lem:gtwisted_startriangles}, one of $\bar{C}_1$ and $\bar{C}_2$ is in $\Delta_1$ and the other one in $\Delta_2$.

		We claim that $v_1$ is not on the boundary of $C_2, C'_1$ or $C'_2$, and at least one of $C'_1$ and $C'_2$, say $C'_1$, has at most one vertex on its boundary.
		Indeed, on the one hand, since $(C_1, C_2)$ is contained in $(\bar{C}_1,\bar{C}_2)$ and $v_1$ is on the boundary of $C_1$, then $v_1$ is not on the boundary of $C_2$ and $v_2$ must be one of $d$ and $c$.
		On the other hand, by Lemma~\ref{lem:two_non_adjacent}, $(C'_1,C'_2)$ must be contained in either $(\bar{C}_1,\bar{C}_2)$ or~$(\bar{C}'_1,\bar{C}'_2)$. Suppose first that $(C'_1,C'_2)$ is contained in $(\bar{C}'_1,\bar{C}'_2)$ (cells marked with red squares in Figure~\ref{fig:Case2v_1}), say $C'_1$ and $\bar{C}'_1$ are in $\Delta_1$, and $C'_2$ and $\bar{C}'_2$ are in $\Delta_2$. Then as $\Delta_1$ and $\Delta_2$ are empty, necessarily $v'_1=e$, $v'_2=c$, $v_1$ is not on the boundary of $C'_1$ or $C'_2$, and both $C'_1$ and $C'_2$ have one vertex on its boundary. Suppose now that $(C'_1,C'_2)$ is contained in $(\bar{C}_1,\bar{C}_2)$, say $C'_1$ in $\bar{C}_1$ and $C'_2$ in $\bar{C}_2$. Again, $v_1$ cannot belong to the boundary of $C'_2$ and $v'_2$ must be either $c$ or $d$. Let us see that in this case of $C_1$ and $C'_1$ both belonging to $\bar{C}_1$, $v_1$ is not on the boundary of $C'_1$ and $C'_1$ only has a vertex on its boundary. As $C_1$ and $C'_1$ are different, there has to be an edge separating $C_1$ and $C'_1$ in $D$.	This edge cannot be adjacent to $v_1$ since $v_1be$ is an empty star triangle at $v_1$ and thus there cannot be any edge emanating from $v_1$ to the empty side of $v_1be$.	Hence, this edge either emanates from $e$ on the empty side of $\Delta_1$ crossing the edge $v_1b$, or crosses $v_1e$ and separates $e$ from $v_1$. In both cases, as $v_1$ belongs to $C_1$, necessarily $v_1$ is not on the boundary of $C'_1$, and $C'_1$ only has a vertex on its boundary (vertex $v'_1=e$). Therefore,  $v_1$ is not on the boundary of $C_2, C'_1$ or $C'_2$, and $C'_1$ has at most one vertex on its boundary, as claimed.
		We remark that in any case, $v_2$ and $v'_2$ are $c$ or $d$.

		We now show that $C_1$ must have two vertices, $v_1$ and $e$, on its boundary. Assume, for a contradiction, that there exists an edge $xy$ crossing $v_1e$ and defining part of the boundary of $C_1$. Note that this edge must separate $C_1$ and $C'_1$ because $v'_1$ is always $e$ (regardless of whether $C'_1$ is in $\bar{C}_1$ or $\bar{C}'_1$), so $v'_1$ is not on the boundary of $C_1$ (nor on the boundary of $C_2$ or $C'_2$ because they are in $\Delta_2$). By Lemma~\ref{lem:gtwisted_triangles_special}, the vertices $x$ and $y$ together with $v_1$ form an empty triangle, which is a star triangle at $x$ and~$y$, containing $C_1$ in its empty side. %
		As $xy$ separates $C_1$ and $C'_1$, the triangle $xyv_1$ cannot contain $C'_1$ on the empty side. Consequently, it has to contain $C'_2$ and thus $v'_2$ has to be one of $x$ and $y$, say $x$. Since $v'_2$ is one of $c$ and $d$, and $xy$ crosses $v_1e$, necessarily $y$ does not belong to $D_0$. Consider the subdrawing $D_6$ induced by the vertices $b, e, v_1, c, d$ and $y$. As the rotation system of $D$ is {\gtwisted}, the rotation system of $D_6$ is also {\gtwisted}, so $D_6$ must be strongly isomorphic to one of the drawings in Figure~\ref{appfig:gtwistedk6}~\cite{plane_gtwisted_journal}. By construction, $C_1$ and $C'_1$ belong to different cells in $D_6$, so $D_6$ has to contain two pairs of antipodal vi-cells by Lemma~\ref{lem:containment}. However, $v_1$ and $v'_1$ belong to only one via-cell, contradicting Observation~\ref{obs:K6}. Thus, $v_1e$ has to be uncrossed in $D$. Consequently, both $v_1$ and $e$ lie on the boundary of~$C_1$.

		Consider now $v'_1$ and the two empty star triangles at $v'_1$ that are not adjacent by hypothesis. By the same arguments as for $v_1$, the cell $C'_1$ has to have two vertices on its boundary.
		This contradicts the observation above that $C'_1$ has only one vertex, $v'_1=e$, on its boundary.
		This completes the proof of Property~\ref{slem:p1}. %
		
		To prove Properties~\ref{slem:p2} and~\ref{slem:p3}, we use that by Property \ref{slem:p1}, there are two empty star triangles at a vertex $v_1$ that are adjacent along an edge $v_1v_2$. Thus, by Lemma~\ref{lem:adjacent_then_maxcross}, this implies that $v_1v_2$ is maximum crossing and the empty star triangles at $v_1$ are also empty star triangles at $v_2$. Besides, by Lemma~\ref{lem:maxcross_then_gtwisted} the via-cells are adjacent to the maximum crossing edge. Therefore, Properties~\ref{slem:p2} and~\ref{slem:p3} hold.
		
		To prove Property~\ref{slem:p4}, assume for a contradiction that there is a third pair $(C''_1, C''_2)$ of antipodal vi-cells. By Properties~\ref{slem:p1}, \ref{slem:p2} and~\ref{slem:p3}, the two pairs $(C_1, C_2)$ and $(C''_1, C''_2)$ define a maximum crossing edge $e$ and the cells $C_1, C_2, C''_1$ and $C''_2$ must by adjacent to $e$. As by Lemma~\ref{lem:only_one_maxcross} any simple drawing has at most one maximum crossing edge, then $e$ must be $v_1v_2$, implying that $(C'_1, C'_2)$ and $(C''_1, C''_2)$ coincide.  Therefore, $D$ cannot have more than two pairs of antipodal vi-cells, so Property~\ref{slem:p4} holds.
\end{proof}

\begin{proof}[Proof of Lemma~\ref{lem:included_cells}]
The containment relations of Lemma~\ref{lem:included_cells} are a direct consequence of Lemma~\ref{lem:containment}. The adjacency to $m$ follows from Lemma~\ref{lem:two_pairs_adjacent}.
\end{proof}

\begin{proof}[Proof of Lemma~\ref{lem:Claim1}]
As the proof of Property~\ref{p:claim1_2} is totally analogous to the proof of Property~\ref{p:claim1_1}, we only prove Property~\ref{p:claim1_1}. For convenience, we rename $\Delta_1 = v_1u_1u'_1$ as $\Delta = v_1uu'$.

We first show that for two grid cells $G^{w_i}$ and $G^{w_j}$ placed before $m$, their intersection is a grid cell $G$ of $D$. By Lemma~\ref{lem:included_cells}, as $C^{w_i}$ and $C^{w_j}$ are before $m$, when removing $w_i$ and $w_j$ there must be a via-cell $C^{w_iw_j}$ in a grid cell $G^{w_iw_j}$ of $D\!\setminus\! \{w_i,\!w_j\}$ before $m$ containing $C^{w_i}$ and $C^{w_j}$. This implies that
the two grid cells $G^{w_i}$ and $G^{w_j}$ must be inside $G^{w_iw_j}$.
In $D\!\setminus\!\{w_i, w_j\}$, the grid cell $G^{w_iw_j}$ is defined by two consecutive edges $r_a$ and $r_b$ around a vertex, so at most $r_{w_i}$ and $r_{w_j}$ can appear between   $r_a$ and $r_b$ in $D$. We distinguish three cases.

Suppose first that none of $r_{w_i}$ and $r_{w_j}$ is between $r_a$ and $r_b$. Thus, $G^{w_iw_j}$ is necessarily an (enlarged) grid cell of $D$, with $r_a$ and $r_b$ being consecutive around $u$, or $u'$ or $v_1$; see Figure~\ref{fig:Gvw_1} for an illustration. Regardless of whether $r_{w_i}$ and $r_{w_j}$ cross $G^{w_iw_j}$ or not, it can be easily checked that $G^{w_i}\cap G^{w_j}$ is a grid cell of $D$. For instance, Figure~\ref{fig:Gvw_1} shows the case that $r_{w_i}$ and $r_{w_j}$ both cross $G^{w_iw_j}$. In this case, $G^{w_i}$ is defined by $r_a, r_b$ and $r_{w_j}$, $G^{w_j}$ is defined by $r_a, r_b$ and $r_{w_i}$, and the intersection of $G^{w_i}$ and $G^{w_j}$ is $G^{w_i}$, which is a grid cell of~$D$.

Suppose now that one of $r_{w_i}$ and $r_{w_j}$, say  $r_{w_i}$, is between $r_a$ and $r_b$; see  Figures~\ref{fig:Gvw_2} and~\ref{fig:Gvw_3}. Note that in this case, $r_a$ and $r_b$ cannot emanate from $v_1$. Then $G^{w_iw_j}$ consists of two (enlarged) grid cells of $D$. Again, regardless of whether $r_{w_j}$ crosses $G^{w_iw_j}$ or not, and whether $G^{w_j}$ is defined by $r_a$ and $r_{w_i}$, or by $r_{w_i}$ and $r_b$, one can check that $G^{w_i}\cap G^{w_j}$ is a grid cell of $D$. For instance, when $r_{w_j}$ crosses $G^{w_iw_j}$ as in Figure~\ref{fig:Gvw_2}, $G^{w_i}$ is defined by $r_a, r_b$ and $r_{w_j}$, $G^{w_j}$ is defined by either $r_a$ and $r_{w_i}$, or by $r_{w_i}$ and $r_{b}$, and the intersection of $G^{w_i}$ and $G^{w_j}$ is a grid cell of $D$.

\begin{figure}[!htb]
	\centering
		\subfloat[]{%
				\includegraphics[scale=0.85,page=1]{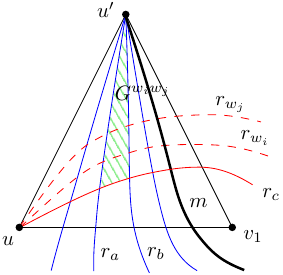}
				\label{fig:Gvw_1}
			}~~~~~~
			\subfloat[]{
				\includegraphics[scale=0.85,page=1]{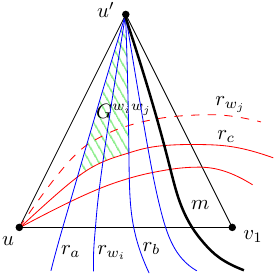}
				\label{fig:Gvw_2}
			}~~~~~~
			\subfloat[]{
				\includegraphics[scale=0.85,page=1]{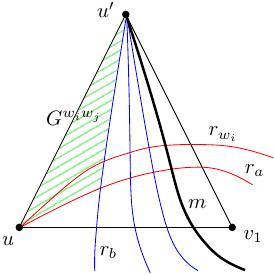}
				\label{fig:Gvw_3}
			}\\
			\subfloat[]{%
					\includegraphics[scale=0.85,page=1]{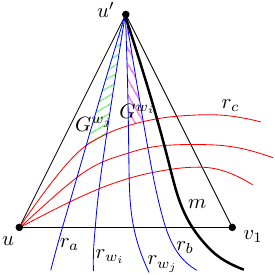}
					\label{fig:Gvw_4}
				}~~~~~~
				\subfloat[]{
					\includegraphics[scale=0.85,page=1]{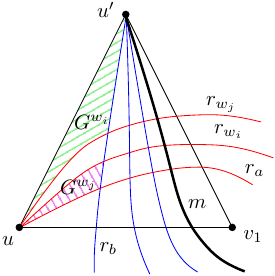}
					\label{fig:Gvw_5}
				}~~~~~~
				\subfloat[]{
					\includegraphics[scale=0.85,page=1]{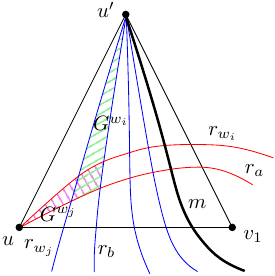}
					\label{fig:Gvw_6}
				}
				\caption{The different possibilities for $G^{w_iw_j}$.}
				\label{fig:Distance2}
			\end{figure}

			Finally, suppose that $r_{w_i}$ and $r_{w_j}$ both are between $r_a$ and $r_b$, say $r_{w_i}$ after $r_a$ counter\-clockwise. There are three possibilities depending on whether the four edges $r_a, r_{w_i}, r_{w_j}$ and $r_b$ emanate from a same vertex (see Figure~\ref{fig:Gvw_4}), three of them emanate from a vertex and the other one from another vertex (see Figure~\ref{fig:Gvw_5}), or two of them emanate from a same vertex and the other two from another vertex (see Figure~\ref{fig:Gvw_6}). In any case, regardless of whether $G^{w_i}$ is defined by $r_a$ and $r_{w_j}$, or by $r_{w_j}$ and $r_b$, and of whether $G^{w_j}$ is defined by $r_a$ and $r_{w_i}$, or by $r_{w_i}$ and $r_b$, the intersection of $G^{w_i}$ and $G^{w_j}$ is a grid cell of $G$, except for the three cases shown in Figures~\ref{fig:Gvw_4}, \ref{fig:Gvw_5} and~\ref{fig:Gvw_6}, where $G^{w_i}$ is defined by $r_{w_j}$ and $r_b$, and $G^{w_j}$ is defined by $r_a$ and $r_{w_i}$.
			
			Let us see that these three last cases cannot happen. Take a vertex $y$ different from $w_i$ and $w_j$ and consider the grid cell $G^{y}$ of ${D\! \setminus\!\{ y\} }$ containing a via-cell $C^y$ of ${D\! \setminus\!\{ y\} }$. We show that $G^y$ cannot be placed anywhere, assuming that $G^{w_i}$ is defined by $r_{w_j}$ and $r_b$, and $G^{w_j}$ is defined by $r_a$ and $r_{w_i}$. Recall that by Lemma~\ref{lem:included_cells}, if $C^y$ is before $m$, then for any $k$, $G^{w_k}$ and $G^y$ must be in a grid cell of $D\!\setminus\! \{w_k,y\}$ after removing $r_{w_k}$ and $r_y$. Using this, we have that $G^{y}$ cannot be placed before $r_{w_i}$, because otherwise $r_{w_j}$ would separate $G^{w_i}$ and $G^{y}$ after removing $r_{w_i}$ and $r_y$. For the same reason, we have that $G^{y}$ cannot be placed after $r_{w_i}$ and before $m$, because otherwise $r_{w_i}$ would separate $G^{w_j}$ and $G^{y}$ after removing $r_{w_j}$ and $r_y$. Besides, $G^{y}$ cannot be placed after $m$. Indeed, if $G^{y}$ is placed after $m$, then by removing $r_{w_j}$ and $r_x$ the cells $C^{w_j}$ and $C^{y}$ are separated by at least $r_{w_i}$ and $m$. By Lemma~\ref{lem:included_cells}, this implies that $C^{w_j}$ should be in a via-cell $C$ of a pair of antipodal vi-cells of $D\!\setminus\!\{w_j,y\}$ and $C^y$ in a via-cell $C'$ of a different pair of antipodal vi-cells of $D\!\setminus\!\{w_j,y\}$. But, when there are two pairs of antipodal vi-cells in a simple drawing, all the via-cells must be adjacent to a common edge by Lemma~\ref{lem:two_pairs_adjacent}, which it is not possible because $r_{w_i}$ and $m$ separate $C$ and $C'$. Therefore, the three cases in Figures~\ref{fig:Gvw_4}, \ref{fig:Gvw_5} and~\ref{fig:Gvw_6} cannot happen and the intersection of $G^{w_i}$ and $G^{w_j}$ must be a grid cell of $D$. 

			We now show that the intersection of $G^{w_1}, G^{w_2}, \ldots , G^{w_s}$ is a common grid cell $G$ of $D$ placed before $m$. Assume, for a contradiction, that $G^{w_i}\cap G^{w_j} = G$, $G^{w_j}\cap G^{w_k} = G'\neq G$, and $G$ is placed before $G'$ in $D$. As $G$ and $G'$ are different and both belong to $G^{w_j}$, then $G^{w_j}$ must consist of the grid cells $G$ and $G'$ of $D$, separated by $r_{w_j}$. This implies that $G^{w_i}$ is on the side of $r_{w_j}$ that contains $G$, and $G^{w_k}$ on the side of $r_{w_j}$ that contains $G'$. But then $G^{w_iw_k}$ should contain $G^{w_i}$ and $G^{w_k}$, which is impossible because $r_{w_j}$ separates $G^{w_i}$ and $G^{w_k}$, a contradiction. Therefore, the intersection of $G^{w_1}, G^{w_2}, \ldots , G^{w_s}$ must a common grid cell $G$ of $D$. This concludes the proof of Property~\ref{p:claim1_1}.

			Finally, to prove Property~\ref{p:claim1_3}, we first show that any grid cell $G^{w_i}$ has to contain $G_c$. Assume for a contradiction that $G^{w_i}$ does not contain $G_c$. Take a grid cell $G^{x_j}$ placed after $m$. Since $G^{w_i}$ (and its corresponding vi-cell $C^{w_i}$) is before $m$ and $G^{x_j}$ (and its corresponding vi-cell $C^{x_j}$) is after $m$, by Lemma~\ref{lem:included_cells} the only possibility to glue such grid cells (vi-cells) in $D\!\setminus\!\{w_i, x_j\}$ (along $m$) is that $D\!\setminus\!\{w_i, x_j\}$ contains two pairs of antipodal vi-cells and the four vi-cells of these two pairs are adjacent to $m$. Thus, $G^{w_ix_j}$ must contain $G_c$ in addition to $G^{w_i}$, and at most $r_{w_i}$ and $r_{x_j}$ can separate $G^{w_i}$ and $G_c$, as $G^{w_i}$ must be glued to $m$ in $D\!\setminus\!\{w_i, x_j\}$. But if we take another grid cell $G^{x_k}$ after $m$, then $G^{w_i}$ should be glued to $m$ in $D\!\setminus\!\{w_i, x_k\}$, which is impossible because $r_{x_j}$ separates $G^{w_i}$ and $G_c$, a contradiction. Therefore, any grid cell $G^{w_i}$ has to contain $G_c$, and the common grid cell $G$ of $D$ in the intersection of all $G^{w_i}$ must be precisely $G_c$. Using a similar argument, the common grid cell $H$ of $D$ in the intersection of all $G^{x_i}$ must be $G_{c+1}$.
\end{proof}

\begin{figure}[!htb]
	\centering
		\subfloat[The edge $w_jw'_j$ separates $u$ and $u'$ ($G^x$ and $G^{w_j}$).]{%
				\includegraphics[scale=0.9,page=1]{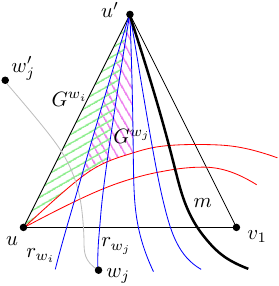}
				\label{fig:Claim3_1}
			}~~~~~~~~~~~~~~~~~~~
			\subfloat[There are edges $w_1w'_1$ and $w_2w'_2$ crossing $G$.]{
				\includegraphics[scale=0.9,page=1]{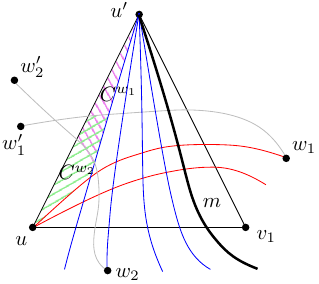}
				\label{fig:Claim3_2}
			}
			\caption{Illustrating the proof of Lemma~\ref{lem:Claim2}.}
			\label{fig:crossing_G1}
		\end{figure}

\begin{proof}[Proof of Lemma~\ref{lem:Claim2}]
Again for convenience, we rename $\Delta_1 = v_1u_1u'_1$ as $\Delta = v_1uu'$.	

We first claim that if two arbitrary grid cells $G^{w_i}$ and $G^{w_j}$ contain a common grid cell $G$ of $D$, then the interior of $C^{w_i}$ (or $C^{w_j}$) and the interior of $G$ cannot be disjoint. Assume that $G$ contains two vertices on its boundary (either $u$ and $u'$, or $u'$ and $v_1$), so $G^{w_i}$ and $G^{w_j}$ as well. Thus, the result is obvious as $C^{w_i}$ and $C^{w_j}$ are incident to at least one of these two vertices. Then assume that $G$ contains only one vertex on its boundary, say $u'$. If $G^{w_i}$ and $G^{w_j}$ only contain $u'$ on the boundary, then the result is again obvious. Hence, the only case left is the case that one of $G^{w_i}$ and $G^{w_j}$, say $G^{w_j}$, has only $u'$ on its boundary (so $C^{w_j}$ is incident to $u'$) and $G^{w_i}$ contains two vertices on its boundary, say $u$ and $u'$; see Figure~\ref{fig:Claim3_1} for an illustration. If $C^{w_i}$ is incident to $u'$, we are done, so we may assume that $C^{w_i}$ is only incident to $u$.  When removing $w_i$ and $w_j$, $C^{w_i}$ and $C^{w_j}$ must be glued in a same via-cell of $D\!\setminus\!\{w_i,\!w_j\}$, so any edge of $D$ crossing $uu'$ and separating $C^{w_i}$ and $C^{w_j}$ must be incident to either $w_i$ or $w_j$. Besides, since $C^{w_i}$ is not incident to $u'$, it is necessary that an edge $w_jw'_j$ with $w'_j\ne w_i$ exists so that $C^{w_i}$ does not contain $u'$ when removing $w_i$.
Consider a vertex $x \notin \{V_0,w_i,w_j,w'_j\}$ (recall that $V_0$ is the set of vertices involved in the two nonadjacent empty star triangles at $v_1$). Since the via-cells $C^{x}$ and $C^{w_i}$ are glued in $D\! \setminus\!{\{w_i,\!x\}}$, the only edges in $D$ between $C^{x}$ and $C^{w_i}$ are edges of $S(x) \cup S(w_i)$ (or $m$ if $D\!\setminus\!\{w_i,\!x\}$ has two pairs of antipodal vi-cells). In particular, $w_jw'_j$ cannot be an edge separating them, so $C^{x}$ cannot contain $u'$. But then, since $G^x$ only contains $u$ on its boundary and $G^{w_j}$ only $u'$, the intersection between $G^x$ and $G^{w_j}$ cannot be a grid cell of $D$, contradicting Lemma~\ref{lem:Claim1}. Therefore, the interior of $C^{w_i}$ ($C^{w_j}$) and the interior of $G$ are not disjoint, as claimed.

We now show that $C^{w_1}, C^{w_2}, \ldots, C^{w_s}$ also contain a common vi-cell $C$ of $D$. By the previous claim, Lemma~\ref{lem:Claim2} is obvious if the common grid cell $G$ has only one vertex on the boundary. Hence, we may assume that $G$ is a grid cell containing two vertices on its boundary, say $u$ and $u'$.
We partition $C^{w_1}, C^{w_2}, \ldots, C^{w_s}$ into three sets $S_1, S_2$ and $S_3$. The set $S_1$ contains the cells $C^{w_i}$ that are only incident to $u'$, the set $S_2$ contains the cells $C^{w_i}$ that are only incident to $u$, and the set $S_3$ contains the cells $C^{w_i}$ that are incident to both $u$ and $u'$. We show that some of $S_1$ and $S_2$ is empty.

Suppose for a contradiction that $S_1$ and $S_2$ are both non-empty. Let $C^{w_1}$ and $C^{w_2}$ be vi-cells in $S_1$ and $S_2$, respectively; see Figure~\ref{fig:Claim3_2} for an illustration.
As explained previously, when removing $w_1$ and $w_2$, $C^{w_1}$ and $C^{w_2}$ must be glued in a same via-cell of $D\!\setminus\!\{w_1,\!w_2\}$, so any edge of $D$ crossing $G$ must be incident to either $w_1$ or $w_2$. Besides, since $C^{w_1}$ and $C^{w_2}$ are different, it is necessary that there exits an edge $w_1w'_1$ ($w'_1\ne w_2$) so that $C^{w_2}$ does not contain $u$ when removing $w_2$, and an edge $w_2w'_2$ ($w'_2\ne w_1$) so that $C^{w_1}$ does not contain $u'$ when removing $w_1$.
Note that $w'_1$ and $w'_2$ can coincide, but do not have to. Consider a vertex $x \notin \{V_0,w_1,w'_1,w_2,w'_2\}$. As before, since the via-cells $C^{x}$ and $C^{w_1}$ are glued in $D\! \setminus\!{\{w_1,\!x\}}$,
$w_2w'_2$ cannot be an edge separating them, so $C^{x}$ cannot contain $u'$. Analogously, $w_1w'_1$ cannot separate $C^{x}$ and~$C^{w_2}$, and $C^x$ cannot contain $u$. Thus, the only possibility for $C^x$ to be a vi-cell is that it is incident to $v_1$. But in such a case, $C^x$ cannot be glued to $C^{w_1}$ by removing $w_1$ and $x$ because $C^x$ is in a grid cell that is after $m$ and $C^{w_1}$ is adjacent to $uu'$. Therefore, $C^{w_1}$ and $C^{w_2}$ must have some vertex in common, contradicting our choice of $S_1$ and $S_2$. As a consequence, at least one of $S_1$ and $S_2$ must be empty, $C^{w_1}, C^{w_2}, \ldots, C^{w_s}$ must share at least one vertex of $G$, and by the previous claim  $C^{w_1}, C^{w_2}, \ldots, C^{w_s}$ must contain a common vi-cell $C$ of $D$.
\end{proof}

\begin{proof}[Proof of Lemma~\ref{claim:proofend}]
	Consider seven vertices $w_1, \ldots , w_7$ not in $V_0$. For $i=1, \ldots , 7$, let $(C_1^{w_i}, C_2^{w_i})$ be a pair of antipodal vi-cells in the subdrawing $D\!\setminus\!\{w_i\}$, and let $G_1^{w_i}$ and $G_2^{w_i}$ be the grid cells in $\Delta_1$ and $\Delta_2$ (of $D\!\setminus\!\{w_i\})$ containing $C_1^{w_i}$ and $C_2^{w_i}$, respectively. From the seven grid cells $G_1^{w_i}$, at least four, say $G_1^{w_1}, G_1^{w_2}, G_1^{w_3}$ and $G_1^{w_4}$, are on a same side of~$m$, say before. We distinguish whether there is a grid cell $G_1^{w_i}$ after $m$ or not.
	
	\paragraph*{Case 1: At least one of the grid cells $G_1^{w_i}$, say $G_1^{w_7}$, is after $m$.}		
	Observe that for the rotation system of $D\! \setminus\!\{ w_1, w_7\} $ to be  \gtwisted, since  $G_1^{w_1}$ (and $C_1^{w_1}$) is before $m$ and $G_1^{w_7}$ (and $C_1^{w_7}$) after $m$, by Corollary~\ref{lem:included_cells} the only option is that $D\! \setminus\! \{ w_1,w_7\}$ has two pairs $(C_1,C_2)$ and $(C'_1, C'_2)$ of antipodal vi-cells. Besides, by Lemma~\ref{lem:two_pairs_adjacent}, $C_1$ and $C'_1$ are the vi-cells in $D\! \setminus\!\{ w_1, w_7\} $ incident to $u_2$ and placed before and after $m$, and $C_2$ and $C'_2$ are the vi-cells in $D\! \setminus\!\{ w_1, w_7\} $ incident to $u'_2$ and placed after and before $m$. In particular, this implies that since $G_1^{w_1}$ is before $m$ in $\Delta_1$, then $G_2^{w_1}$ must be after $m$ in $\Delta_2$. For the same reason, $G_2^{w_2}, G_2^{w_3}$ and $G_2^{w_4}$ are after $m$ in $\Delta_2$. By Lemmas~\ref{lem:Claim1} and~\ref{lem:Claim2}, there is a vi-cell $C$ in $\Delta_1$ contained in $C_1^{w_1}$, $C_1^{w_2}, C_1^{w_3}$ and $C_1^{w_4}$, and a vi-cell $C'$ in $\Delta_2$ contained in $C_2^{w_1}$, $C_2^{w_2}, C_2^{w_3}$ and $C_2^{w_4}$. Let us see that $(C, C')$ are antipodal in $D$. As the pairs of antipodal vi-cells  $(C_1^{w_1}, C_2^{w_1}), (C_1^{w_2}, C_2^{w_2})$, $(C_1^{w_3}, C_2^{w_3})$ and $(C_1^{w_4}, C_2^{w_4})$ contain $(C,C')$, then for any triangle $y_1y_2y_3$ in $D$ such that $w_1$, $w_2$, $w_3$ or $w_4$ are not in $\{y_1,y_2,y_3\}$, the cell $C$ lies on one side of $y_1y_2y_3$ and the cell $C'$ in the other.  Since at least one of $\{w_1,w_2,w_3,w_4\}$ has to be different from $\{y_1,y_2,y_3\}$, this implies that $C$ and $C'$ are antipodal in $D$. Therefore, $(C,C')$ is a pair of antipodal vi-cells in $D$.
	
	\paragraph*{Case 2: None of the grid cells $G_1^{w_i}$ is after $m$.}
	This means, all of them are before $m$. Consider the grid cells $G_2^{w_i}$ in $\Delta_2$. If there were grid cells before and after $m$ in $\Delta_2$, we could argue as in the previous paragraph, interchanging the roles of $\Delta_1$ and $\Delta_2$. Thus, we may assume that all the grid cells $G_2^{w_i}$ are on one of the sides of $m$. Again, by Lemmas~\ref{lem:Claim1} and~\ref{lem:Claim2}, there is a vi-cell $C$ in $\Delta_1$ contained in all the vi-cells $C_1^{w_i}$, and a vi-cell $C'$ in $\Delta_2$ contained in all the vi-cells $C_2^{w_i}$. Given a triangle~$y_1y_2y_3$, as at least one of $\{w_1, \ldots , w_7\}$ has to be different from $\{y_1,y_2,y_3\}$, we have that $(C,C')$ is a pair of antipodal vi-cells in $D$.
\end{proof}

\section{Missing proofs of Section~\ref{sec:bipartition}}

\begin{proof}[Proof of Lemma~\ref{lem:compatible}]

\begin{figure}[!ht]
	\centering
	\includegraphics[scale=0.53,page=1]{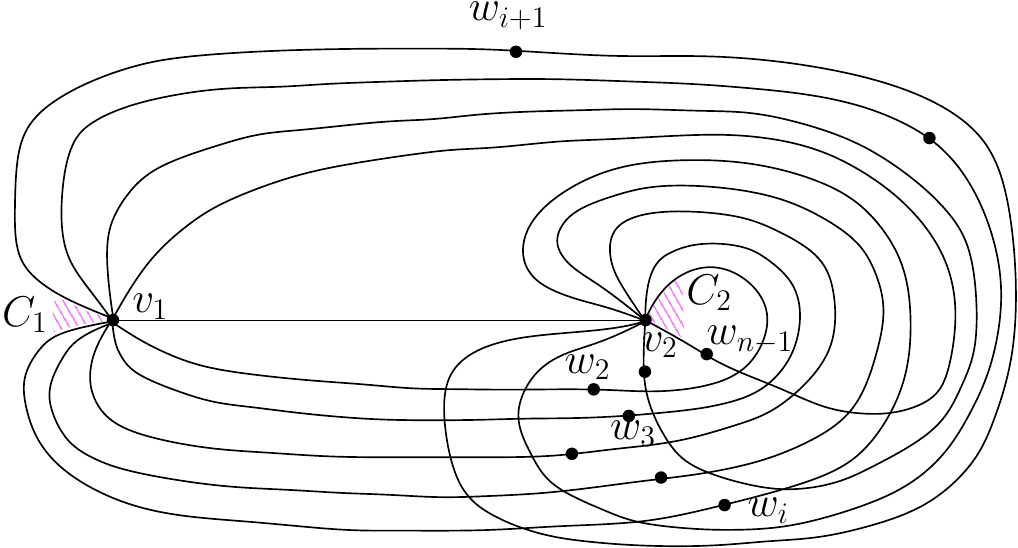}
	\caption{The rotations of $v_1$ and $v_2$.}
	\label{fig:approtations}
\end{figure}

\begin{figure}[!htb]
	\centering
		\subfloat[]{%
				\includegraphics[scale=0.53,page=1]{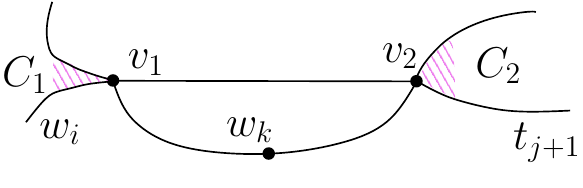}
				\label{app:fig:lemmaw1}
			}~~~~~~~~
			\subfloat[]{
				\includegraphics[scale=0.53,page=1]{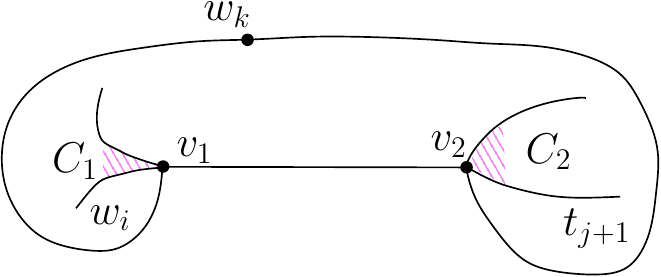}
				\label{app:fig:lemmaw2}
			}
			\caption{If $w_k$ appears before $C_1$ around $v_1$, then $w_k$ cannot appear after $C_2$ around $v_2$.}
			\label{app:fig:lemmaw}
		\end{figure}

		\begin{figure}[!htb]
			\centering
				\subfloat[]{%
						\includegraphics[scale=0.53,page=1]{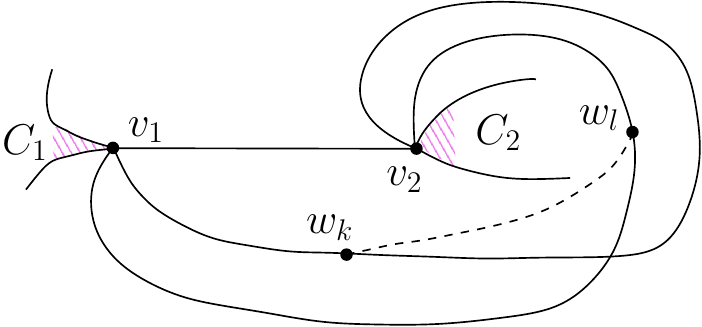}
						\label{app:fig:lemmaa}
					}~~~~~~~~
					\subfloat[]{
						\includegraphics[scale=0.53,page=1]{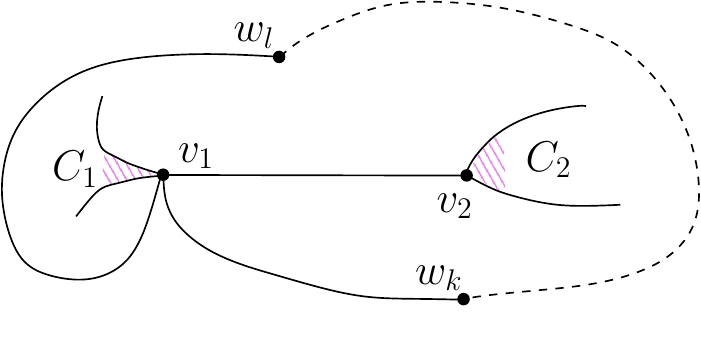}
						\label{app:fig:lemmab}
					}
					\caption{If $w_k$ appears before $w_l$ around $v_1$, then $w_k$ cannot appear before $w_l$ around $v_2$.}
					\label{app:fig:lemma}
				\end{figure}

				We start by proving~\textsf{(i)}.
				Figure~\ref{fig:approtations} shows an example of the rotations of $v_1$ and $v_2$.

				Assume that the (clockwise) rotation of $v_2$ is $\{v_1=t_1, t_2, \ldots, t_{n-1}\}$ and that the edges $v_2t_j$ and $v_2t_{j+1}$, for some $j\in \{1,\ldots, n-1\}$, define part of the boundary of $C_2$.

				We first show by contradiction that if a vertex appears before $C_1$ around $v_1$ (clockwise from $v_2$), then it also appears before $C_2$ around $v_2$ (clockwise from $v_1$), that is, a vertex $w_k$ with $2\le k \le i$ corresponds to a vertex $t_{k'}$ with $2 \le k' \le j$. Suppose to the contrary that a vertex $w_k$ with $2\le k \le i$ corresponds to a vertex $t_{k'}$ with $j+1 \le k' \le n-1$ (see Figure~\ref{app:fig:lemmaw} for a depiction). When rotating clockwise around $v_1$ from $v_2$, we can be either inside or outside the triangle $v_1v_2w_k$, depending on how the edges $v_1v_2$, $v_1w_k$, and $w_kv_2$ are drawn in $D$. Since we are supposing that $j+1 \le k' \le n-1$, if we are inside $v_1v_2w_k$ (Figure~\ref{app:fig:lemmaw1}), then $C_1$ and $C_2$ are necessarily outside $v_1v_2w_k$, and if we are outside $v_1v_2w_k$ (Figure~\ref{app:fig:lemmaw2}), then $C_1$ and $C_2$ are necessarily inside $v_1v_2w_k$. In any case, this contradicts that $C_1$ and $C_2$ are antipodal. Therefore, a vertex $w_k$ with $2\le k \le i$ corresponds to a vertex $t_{k'}$ with $2 \le k' \le j$.
				
				Using a similar reasoning, we also have that if a vertex appears after $C_1$ around $v_1$ (clockwise from $v_2$), then it appears after $C_2$ around $v_2$ (clockwise from $v_1$), that is, a vertex $w_k$ with $i+1\le k \le n-1$ corresponds to a vertex $t_{k'}$ with $j+1 \le k' \le n-1$. Hence, in the clockwise rotation of $v_2$ beginning at $v_1$, vertices $w_k$ with $k \le i$ appear first and then vertices $w_k$ with $k \ge i+1$.

				We now show, again by contradiction, that for two vertices $w_k$ and $w_l$ with $2\le k < l \le i$, $w_l$ appears before $w_k$ in the (clockwise) rotation of $v_2$ from $v_1$. Suppose to the contrary that $w_k$ appears before $w_l$. Then the edges $v_1w_l$ and $v_2w_l$ emanate in different sides of the triangle $v_1v_2w_k$ from $v_1$ and $v_2$, respectively. Thus, necessarily either $v_1w_l$ and $w_kv_2$ cross or $v_1w_k$ and $w_lv_2$ cross (see Figure~\ref{app:fig:lemmaa} for a depiction). Besides, since there is exactly one crossing in the subdrawing induced by any four vertices of a {\gtwisted} drawing, $w_kw_l$ cannot cross any of the edges $v_1v_2, v_1w_k, v_1w_l, v_2w_k$ and $v_2w_l$, because either $v_1w_l$ and $w_kv_2$ or $v_1w_k$ and $w_lv_2$ cross (see Figure~\ref{app:fig:lemmaa}). However, as $w_k$ and $w_l$ appear before $C_1$ around $v_1$ (clockwise from $v_2$), if $w_kw_l$ does not cross $v_1v_2$, then $C_1$ and $C_2$ cannot be in different sides of the triangle $v_1w_kw_l$ (see Figure~\ref{app:fig:lemmab}), contradicting that $C_1$ and $C_2$ are antipodal. Therefore, for two vertices $w_k$ and $w_l$ with $2\le k < l \le i$, $w_l$ appears before $w_k$ in the (clockwise) rotation of $v_2$ beginning at $v_1$.
				
				In an analogous way, we can also prove that for two vertices $w_k$ and $w_l$ with $i+1\le k < l \le n-1$, then $w_l$ appears before $w_k$ in the (clockwise) rotation of $v_2$ beginning at $v_1$.
				
				As a consequence, since the vertices before $C_1$ around $v_1$ (beginning at $v_2$) must appear before $C_2$ around $v_2$ (beginning at $v_1$) in opposite order, and the vertices after $C_1$ around $v_1$ must appear after $C_2$ around $v_2$ in opposite order, the (clockwise) rotation of $v_2$ must be $\{v_1, w_i, w_{i-1}, \ldots , w_2,w_{n-1},$ $w_{n-2},$ $\ldots , w_{i+1}\}$ and the edges $v_2w_2$ and $v_2w_{n-1}$ define part of the boundary of $C_2$, which completes the proof of~\textsf{(i)}.
				
				The proofs of~\textsf{(ii)} and~\textsf{(iii)} are analogous, except that around $v_1$, there are no edges $v_1w_j$ ($2 \le j \le n-1$)  before (Case~\textsf{(ii)}) or after (Case~\textsf{(iii)}) the cell $C_1$, respectively. Hence, one of the groups $\{w_2,\ldots,w_i\}$ and $\{w_{i-1},\ldots,w_{n-1}\}$ is empty, while the other one contains all vertices $\{w_2,\ldots,w_{n-1}\}$. In both cases, this yields a clockwise rotation of $\{v_1, w_{n-1}, w_{n-2}, \ldots , w_{2}\}$ at $v_2$. In Case~\textsf{(ii)} (no edge before $C_1$), the edges $v_2v_1$ and $v_2w_{n-1}$ are on the boundary of $C_2$, while in Case~\textsf{(iii)} (no edge after $C_1$), the edges  $v_2v_1$ and $v_1w_{2}$ are on the boundary of $C_2$.
\end{proof}

\begin{proof}[Proof of Theorem~\ref{the:bipartition}]
	
		$\Rightarrow$ We first show that if $D$ is {\gtwisted}, then there exists a bipartition of the vertices satisfying \textsf{(i)}--\textsf{(v)}.

	\begin{figure}[!htb]
		\centering
			\subfloat[]{%
					\includegraphics[scale=0.53,page=1]{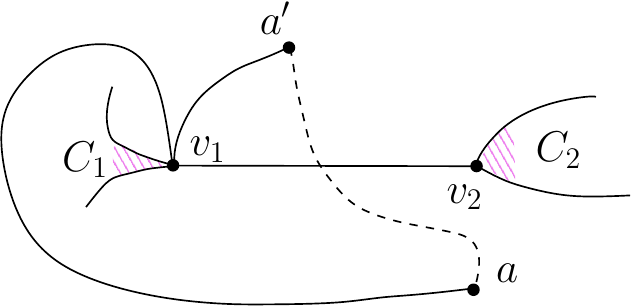}
					\label{fig:apprr}
				}~~~~~~~~~~~~~
				\subfloat[]{
					\includegraphics[scale=0.53,page=1]{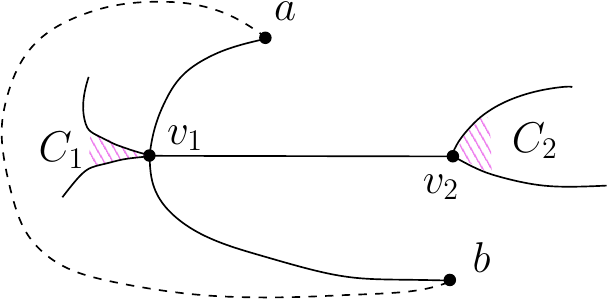}
					\label{fig:apptb}
				}
				\caption{Illustrating the proof of \textsf{(i)} and \textsf{(iii)} in Theorem~\ref{the:bipartition}.}\label{fig:appiii}
			\end{figure}

			Since $D$ is {\gtwisted}, there exists a pair $(C_1, C_2)$ of antipodal vi-cells. Let $v_1$ be a vertex of $V$ incident to $C_1$ and let $v_2\ne v_1$ be a vertex of $V$ incident to $C_2$. Suppose that $\{v_2=w_1, w_2, \ldots ,  w_{n-1}\}$ is the (clockwise) rotation of $v_1$ such that $v_1w_i$ and $v_1w_{i+1}$ define part of the boundary of $C_1$, for some $i\in \{1,\ldots, n-1\}$. (Note that $A$ or $B$ might be empty, then any statement on vertices in that empty set are vacuously true.)
			
			We define a set $A$ and a set $B$ using Lemma~\ref{lem:compatible} depending on the value of $i$ and will show for that bipartition that it satisfies \textsf{(i)}--\textsf{(iv)}. If $i\ne 1,n-1$, by Lemma~\ref{lem:compatible}, the (clockwise) rotation of $v_2$ is $\{v_1, w_i, w_{i-1}, \ldots ,$ $w_2,w_{n-1}, w_{n-2},$ $\ldots , w_{i+1}\}$ and $v_2w_2$ and $v_2w_{n-1}$ define part of the boundary of~$C_2$. In this case, we take $B:=\{w_2, w_3, \ldots , w_i\}$ and $A:= \{w_{i+1}, w_{i+2}, \ldots , w_{n-1}\}$. If $i=1$, again by Lemma~\ref{lem:compatible}, the (clockwise) rotation of $v_2$ is $\{v_1, w_{n-1}, w_{n-2}, \ldots , w_{2}\}$ and the edges $v_2v_1$ and $v_2w_{n-1}$ define part of the boundary of $C_2$. In this case, we take $A:= \{v_1, w_{n-1}, w_{n-2}, \ldots , w_{2}\}$ and $B := \emptyset$. Finally, if $i=n-1$, by Lemma~\ref{lem:compatible}, the (clockwise) rotation of $v_2$ is $\{v_1, w_{n-1}, w_{n-2}, \ldots , w_{2}\}$ and the edges $v_2v_1$ and $v_1w_{2}$ define part of the boundary of $C_2$. In this case, we take $A:= \emptyset$ and $B := \{v_1, w_{n-1}, w_{n-2}, \ldots , w_{2}\}$.
			
			To prove that $A$ and $B$ induce a partition fulfilling \textsf{(i)}--\textsf{(v)}, we first show \textsf{(i)} and \textsf{(ii)} hold. If $A$ contains at most one vertex, \textsf{(i)} is vacuously true. Otherwise, take two vertices $a$ and $a'$ in $A$.
			Since both appear after $C_1$ in the rotation of $v_1$ beginning at $v_2$, if $aa'$ does not cross $v_1v_2$, then $C_1$ and $C_2$ are on the same side of the triangle $v_1 a a'$ (see Figure~\ref{fig:apprr} for an example). This contradicts that $C_1$ and $C_2$ are antipodal cells. Hence, \textsf{(i)} follows.
			By analogous arguments for $B$, \textsf{(ii)}  follows.
			
			We now show \textsf{(iii)} holds for $A$ and $B$. If $A=\emptyset$ or $B=\emptyset$, \textsf{(iii)} is vacuously true. Otherwise, take a vertex $a\in A$ and a vertex $b\in B$. Since $b$ is before $C_1$ and $a$ is after $C_1$ around $v_1$, the edge $ab$ cannot cross $v_1v_2$ so that $C_1$ and $C_2$ are in different sides of the triangle $v_1ab$ (see Figure~\ref{fig:apptb} for an example). Hence, \textsf{(iii)} follows.
			
			Finally, \textsf{(iv)} and \textsf{(v)} follow directly from the definitions of $A$ and $B$. Therefore, $A$ and $B$ satisfy \textsf{(i)}--\textsf{(v)}.
			
			$\Leftarrow$ We now prove that if there exist two vertices $v_1$ and $v_2$ and a bipartition $A\cup B$ of the vertices in $V\!\setminus\!\{v_1,v_2\}$ satisfying  \textsf{(i)}--\textsf{(v)}, then $D$ is {\gtwisted}.
			Let $\{v_2, b_1, \ldots, b_{n_1},$ $a_1, \ldots , a_{n_2}\}$ be the (clockwise) rotation of $v_1$ beginning at $v_2$, and let $\{v_1, b'_1, \ldots, b'_{n_1},$ $a'_1, \ldots , a'_{n_2}\}$ be the (clockwise) rotation of $v_2$ beginning at $v_1$. Let $C_1$ be the vi-cell of $D$ defined by the edges $v_1b_{n_1}$ and $v_1a_1$, and let $C_2$ be the vi-cell of $D$ defined by the edges $v_2b'_{n_1}$ and $v_2a'_1$. We show that $C_1$ and $C_2$ are antipodal by analyzing all possible triangles in~$D$.

			We first take three vertices $x,y,z$ different from $v_1$ and $v_2$. Suppose first that all of them belong to the same subset of the bipartition, say $B$. By \textsf{(ii)}, $v_1v_2$ crosses the three edges of the triangle $xyz$. Hence, since $v_1v_2$ crosses the boundary of $xyz$ three times, the vertices $v_1$ and $v_2$ (and hence the cells $C_1$ and $C_2$) are necessarily in different sides of $xyz$ (see Figure~\ref{fig:appbbb}). Suppose now that only two of $x$, $y$, and $z$ belong to the same subset of the bipartition, say $B$. By \textsf{(ii)} and \textsf{(iii)}, $v_1v_2$ crosses the boundary of $xyz$ only once, implying again that $v_1$ and $v_2$ (and hence $C_1$ and $C_2$) are in different sides of $ xyz$ (see Figure~\ref{fig:appbbr}).

			\begin{figure}[!htb]
				\centering
					\subfloat[$x$, $y$, and $z$
					belong to the same set.]{%
							\includegraphics[scale=0.54,page=1]{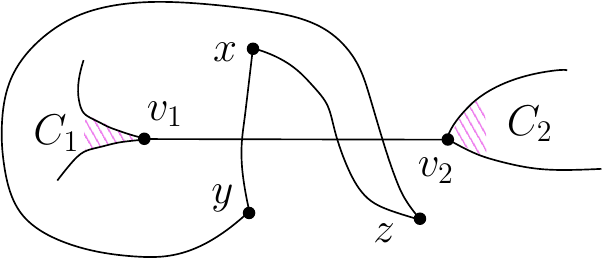}
							\label{fig:appbbb}
						}~~~~~~~~~
						\subfloat[$z$ is in a different set than $x$ and $y$.]{
							\includegraphics[scale=0.54,page=1]{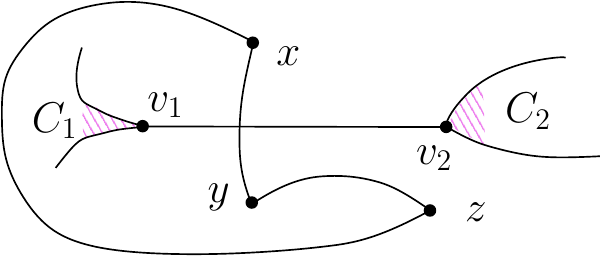}
							\label{fig:appbbr}
						}
						\caption{$x,y$ and $z$ are different from $v_1$ and $v_2$. $C_1$ and $C_2$ are on different sides of $\Delta xyz$.} %
						\label{fig:appthree}
					\end{figure}

					\begin{figure}[!htb]
						\centering
							\subfloat[$x$ and $y$ belong to the
							same set.]{%
									\includegraphics[scale=0.54,page=1]{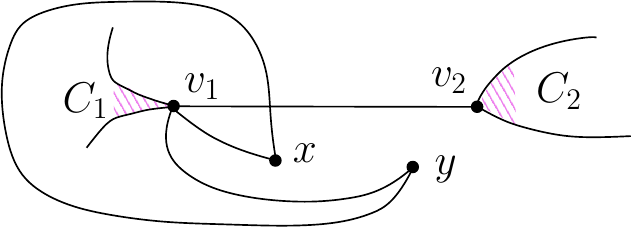}
									\label{fig:appubbi}
								}~~~~~~~~~
								\subfloat[$x$ and $y$ belong to different sets.]{
									\includegraphics[scale=0.54,page=1]{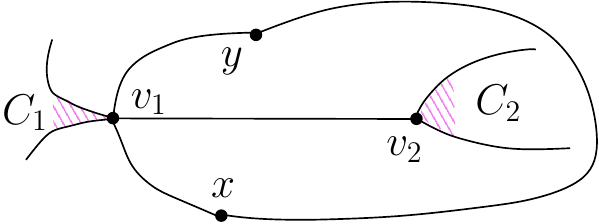}
									\label{fig:appubri}
								}
								\caption{$x$ and $y$ are different from $v_2$. $C_1$ and $C_2$ are on different sides of $\Delta v_1xy$.}%
							\label{fig:apptwo}
						\end{figure}

						\begin{figure}[!htb]
							\centering
								\subfloat[]{%
										\includegraphics[scale=0.54,page=1]{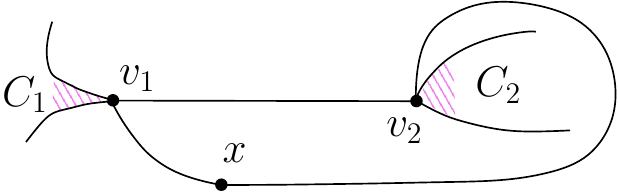}
										\label{fig:appuvbi}
									}~~~~~~~~~
									\subfloat[]{
										\includegraphics[scale=0.54,page=1]{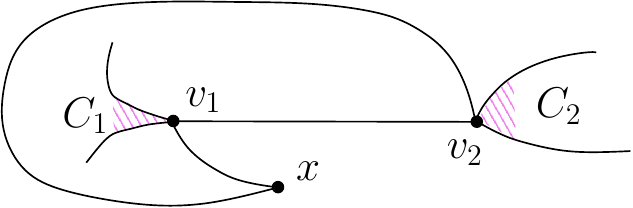}
										\label{fig:appuvbo}
									}
									\caption{$C_1$ and $C_2$ are on different sides of $\Delta v_1v_2x$.}\label{fig:appone}
								\end{figure}

								We now take one of $v_1$ and $v_2$, say $v_1$, and two vertices $x,y$ different from $v_1$ and~$v_2$. Suppose first that $x$ and $y$ belong to the same subset of the bipartition, say $B$, so $v_1v_2$ crosses $xy$ by \textsf{(ii)}. Besides, by the definition of $C_1$, the vertices $x$ and $y$ appear before $C_1$ when rotating clockwise around $v_1$ beginning at $v_2$. These two facts imply that if $v_1v_2$ is inside the triangle $v_1xy$ when emanating from $v_1$, then $C_1$ is also inside $v_1xy$ (see Figure~\ref{fig:appubbi}), and if $v_1v_2$ is outside $v_1xy$ when emanating from $v_1$, then $C_1$ is also outside $v_1xy$. %
								Thus, since $v_1v_2$ crosses $xy$, then $v_2$ (and $C_2$) must be outside $v_1xy$ if $C_1$ is inside $v_1xy$ (see Figure~\ref{fig:appubbi}), and $v_2$ (and $C_2$) must be inside $v_1xy$ if $C_1$ is outside $v_1xy$.
								
								Suppose now that $x\in B$ and $y\in A$ belong to different subsets of the bipartition, so $v_1v_2$ does not cross $xy$ by \textsf{(iii)}. By the definition of $C_1$, $x$ appears before $C_1$ and $y$ after $C_1$ around $v_1$, when rotating clockwise from $v_2$. Hence, if $v_1v_2$ is inside the triangle $v_1xy$ when emanating from $v_1$, then $C_1$ is outside $v_1xy$ (see Figure~\ref{fig:appubri}), and if $v_1v_2$ is outside $v_1xy$ when emanating from $v_1$, then $C_1$ is inside $v_1xy$. %
								Since $v_1v_2$ does not cross $xy$, then $v_2$ (and $C_2$) must be inside $v_1xy$ if $C_1$ is outside $v_1xy$ (see Figure~\ref{fig:appubri}), and $v_2$ (and $C_2$) must be outside $v_1xy$ if $C_1$ is inside $v_1xy$.

								Finally, we take $v_1, v_2$ and an arbitrary vertex $x$. Without loss of generality, we may assume that $x$ belongs to $B$, so $x$ appears before $C_1$ when rotating clockwise around $v_1$ from~$v_2$, and before $C_2$ when rotating clockwise around $v_2$ from $v_1$. Then necessarily $C_1$ and $C_2$ must be in different sides of the triangle $v_1v_2x$ (see Figure~\ref{fig:appone}).
								
								Therefore, since for every triangle of $D$ the vi-cells $C_1$ and $C_2$ are in different sides, then they are antipodal and $D$ is {\gtwisted}.
\end{proof}

\section{Missing proofs of Section~\ref{sec:algo}}

\begin{proof}[Proof of Lemma~\ref{lem:algo:step1}]
	Whether a triangle $xyz$ is a star triangle at~$x$ is determined by whether there exists a vertex $w$ such that $xw$ crosses~$yz$. 	
	From the rotation system, deciding if two edges cross or not requires constant time, thus deciding whether $xyz$ is a star triangle at~$x$ requires linear time by checking for every vertex $w$ if the edges $xw$ and $yz$ cross or not. By Lemma~\ref{lem:gen_startriangles}, a star triangle $xyz$ at $x$ is empty if and only if $y$ and $z$ are consecutive in the rotation of $x$. Thus, to find the empty star triangles at $x$, we only need to explore the (linearly many) pairs of vertices $y,z$ that are consecutive in the rotation of~$x$. Consequently, computing all empty star triangles at~$x$ from the rotation $R$ can be done in $O(n^2)$ time.
\end{proof}

\begin{proof}[Proof of Lemma~\ref{lem:algo:step2}]
	By Lemma~\ref{lem:compatible}, in a generalized twisted drawing there is always a pair $(v_1,v_2)$ of compatible vertices. Further, if $(C_1, C_2)$ is a pair of antipodal vi-cells, then $v_1$ is on the boundary of $C_1$ and $v_2$ on the boundary of $C_2$ (or vice versa).
	
	We look for compatible vertices using the previously calculated empty triangles. Let $x$ be a vertex. If there are not exactly two empty star triangles at $x$, then $R$ is not generalized twisted by Lemma~\ref{lem:gtwisted_startriangles}. Otherwise, we denote by $\Delta = xyz$ and $\Delta' = xy'z'$ the two empty star triangles at~$x$. (We remark that one of $y$ and $z$ might coincide with one of $y'$ and $z'$). If a simple drawing of $K_n$ has a pair of antipodal vi-cells  $(C'_1, C'_2)$, then by Lemma~\ref{lem:gtwisted_startriangles}, $C'_1$ ($C_2'$) must be in $\Delta$ ($\Delta'$), and a vertex of $\Delta$ ($\Delta'$) is incident to $C'_1$
	($C'_2$). Thus, to search for compatible vertices, we only need to check pairs $(v_1,v_2)$ of vertices in the set $S=\{(y,y'), (y,z'), (y,x), (z,y'), (z,z'),$ $(z,x), (x,y'), (x,z')\}$.
	
	Given a pair $(v_1,v_2)$ in $S$, suppose that the rotations of $v_1$ and $v_2$ in $R$ are $\{v_2=w_1, w_2, \ldots ,  w_{n-1}\}$ and $\{v_1=t_1, t_2, \ldots ,  t_{n-1}\}$, respectively. The algorithm can check in linear time if $v_1$ and $v_2$ are compatible as follows. It explores the vertices $w_2, w_3, \ldots $ in order until finding a vertex $w_i$ such that $w_i = t_2$, for some $i$. Then it explores the vertices $t_2, t_3, \ldots , t_i$ in order, checking if this order corresponds to the order $w_i, \ldots , w_2$. If so, the algorithm checks if the order $t_{i+1}, \ldots , t_{n-1}$ is $w_{n-1}, \ldots , w_{i+1}$. %
	This process of checking the rotations requires linear time. %
	
	When there is a pair $(v_1,v_2)$ of compatible vertices, the algorithm checks in $O(n^2)$ time if the properties \textsf{(i)}--\textsf{(v)} of Theorem~\ref{the:bipartition} are satisfied for this pair $(v_1,v_2)$. Let $\{v=w_1, w_2, \ldots ,  w_{n-1}\}$ be the rotation of $v_1$. We define $A$ and $B$ as follows. If the rotation of $v_2$ is $\{v_1, w_{n-1}, \ldots , w_{2}\}$, then $B = \{w_2, \ldots , w_{n-1}\}$ and $A=\emptyset$ (or $A = \{w_2, \ldots , w_{n-1}\}$ and $B=\emptyset$), and if the rotation of $v_2$ is $\{v_1, w_i, \ldots , w_2,w_{n-1}, \ldots , w_{i+1}\}$, for some $i$ different from $1$ and $n-1$, then $B=\{w_2, \ldots , w_i\}$ and $A=\{w_{i+1}, \ldots , w_{n-1}\}$.
	With these definitions, $A$ and $B$ clearly satisfy \textsf{(iv)} and \textsf{(v)} of Theorem~\ref{the:bipartition}.
	Now, the algorithm checks if \textsf{(i)}, \textsf{(ii)}, and \textsf{(iii)} of Theorem~\ref{the:bipartition} are satisfied, that is, if for any pair of vertices $a$ and $a'$ in $A$, the edge $aa'$ crosses $v_1v_2$, if for any pair of vertices $b$ and $b'$ in $B$, the edge $bb'$ crosses $v_1v_2$, and if for any pair of vertices $a\in A$ and $b\in B$, the edge $ab$ does not cross $v_1v_2$.
	As testing if two edges cross requires constant time, checking if \textsf{(i)}--\textsf{(v)} are satisfied for a pair $(v_1,v_2)$ requires $O(n^2)$ time.
	
	Therefore, since $|S|$ is a constant, deciding if there is a pair $(v_1,v_2)$ in $S$ such that $v_1$ and $v_2$ are compatible and the properties \textsf{(i)}--\textsf{(v)} of Theorem~\ref{the:bipartition} are satisfied for this pair can be done in $O(n^2)$ time.	
\end{proof}
\end{document}